\documentclass[%
 reprint,
nofootinbib,
 amsmath,amssymb,
 aps,
prstab,
]{revtex4-2}
\usepackage[caption=false]{subfig}
\usepackage{diagbox}
\usepackage{siunitx}
\usepackage{tikz}

\usepackage{graphicx}
\usepackage{dcolumn}
\usepackage{bm}

\numberwithin{equation}{section}

\begin{document}

\preprint{APS/123-QED}

\title{Controlling the transverse multipole components in rf cavity modes using the azimuthal modulation method}%

\author{L.~M.~Wroe}
 \email{laurence.wroe@cern.ch}
\author{W.~Wuensch}%
\affiliation{%
 CERN, CH-1211 Geneva-23, Switzerland
}%


\author{R.~J.~Apsimon}
\affiliation{
 Engineering Department, Lancaster University, LA1 4YW, United Kingdom} \affiliation{Cockcroft Institute,
Daresbury Laboratory, Warrington, WA4 4AD, United Kingdom}


\begin{abstract}
    Recent work introduced a systematic method for designing so-called azimuthally modulated rf cavities that support transverse magnetic modes composed of user-desired multipoles, enabling precision control of the magnitude and orientation of multipolar components in rf cavity design. This paper extends this method to practical implementation by deriving the multipolar expansion of the longitudinal electric field in such rf cavities with beam pipes, as well as the momentum change of ultra-relativistic particles traversing these modes. The derived equations explicitly show the radial variation of the change in longitudinal and transverse momentum follows a polynomial rather than Bessel-function relationship. The expression for the longitudinal electric field is then compared to a field map obtained from the 3D electromagnetic simulation of an azimuthally modulated cavity designed to support a mode composed of monopole, dipole, and quadrupole components. Beam dynamics studies are presented to assess the derived expressions for the change in momentum, including the effects of relaxing the ultra-relativistic assumption. Finally, two example applications are presented: the first demonstrates the removal of unwanted transverse multipoles to create a multipole-free accelerating structure with a single-port coupler, whereas the second illustrates the synthesis of desired multipoles to create an rf cavity that transforms the transverse  distribution of a beam from Gaussian to uniform.


\end{abstract}

\maketitle

\section{\label{sec:level1}Introduction}

RF cavities in modern particle accelerators typically operate in transverse magnetic TM$_{m10}$ modes, where the integer $m$ denotes the azimuthal variation (or multipolar order) of the mode. The monopolar $m=0$ TM$_{010}$ mode is widely used for accelerating charged particles across various energy scales and gradients. This mode also finds applications in deceleration~\cite{DecelerationCLIC,DecelerationERL}, linearisation~\cite{Linearisation1,Lineariser2}, velocity bunching~\cite{VelocityBunching}, and bunch length and energy compression~\cite{BunchCompression,EnergyCompression}. RF cavities operating in transverse $m\geq1$ modes also have dedicated applications. For example, the dipolar $m~=~1$ TM$_{110}$ mode is used to deflect and separate a particle beam into multiple beams~\cite{DeflectorBrookhaven,KaonSeparator}, control electron beams in free-electron lasers through incorporation in emittance exchangers~\cite{EmittanceExchange1,Emittance2}, compress X-ray pulses in synchrotron light sources~\cite{ShortXray1,ShortXray2}, and exploit a position-dependent transverse kick to enable measurement of longitudinal beam parameters, such as bunch length~\cite{DeflectingTsinhua,Polarix}, or to enable luminosity control in particle colliders as crab cavities~\cite{Crab1,Crab2}. The quadrupolar $m~=~2$ TM$_{210}$ mode is another mode that finds application in Landau damping transverse beam instabilities~\cite{LandauDamping,Landau2}.

TM$_{m10}$ modes are supported by pillbox cavities with a circular cross-section, with modes of distinct $m$ resonating at different frequencies for a given cavity radius. Previous research introduced the azimuthally modulated method (AMM) for designing so-called azimuthally modulated rf cavities that support TM$_{\{m\}10}$ modes, where $\{m\}$ denotes the set of multipolar orders that comprise the mode~\cite{WroeAMM}. The AMM offers a method for controlling the individual strengths and orientations of the multipolar components of the longitudinal electric field, thereby providing a means of precisely tailoring the field for dedicated and specific applications. 

This paper makes significant advancements to the AMM, enabling its practical use in the design of rf cavities. These advancements are underpinned by derivations presented in Sec.~\ref{sec:Theory} which describe the multipolar form of the longitudinal electric field in azimuthally modulated cavities with beam pipes that support TM$_{\{m\}10}$ modes, as well as the longitudinal and transverse momentum change of rigid, ultra-relativistic, parallel particles that traverse these modes. An evaluation and benchmarking of these equations against 3D electromagnetic simulations and beam dynamics studies is then presented in  Sec.~\ref{sec:Analysis}. First, the multipolar form of the longitudinal electric field in an azimuthally modulated cavity designed to support a TM$_{\{0,1,2\}10}$ mode is compared to the field map obtained from 3D simulations of the cavity. Second, the multipolar form of the longitudinal and transverse momentum changes in the same cavity is compared to beam dynamics simulations that track particles through the calculated field maps. Additionally, the effects of relaxing the assumptions of rigid, ultra-relativistic particles are investigated. 

Two example applications of the AMM are presented to demonstrate its potential in rf cavity design. In Sec.~\ref{sec:MultipoleFree}, the AMM is used to remove unwanted transverse multipoles introduced by incorporating a slot-based power coupler into an accelerating structure. These unwanted multipoles cause transverse kicks which in turn lead to emittance growth~\cite{EmittanceGrowth}. To address this, structures have been made that incorporate 2-port, dual-feed power couplers to remove the dipole component and a racetrack cavity design to eliminate the quadrupole component~\cite{DualFeedCoupler}. Other structures incorporate 4-port, quad-feed power couplers to mitigate both dipole and quadrupole components simultaneously~\cite{QuadFeed1,QuadFeed2}. We demonstrate how the AMM can mitigate the transverse multipoles up to any order to negligible levels with just a single-port to create a multipole-free accelerating structure. This approach can minimise the cost, design complexity, and manufacturing time of rf structures, as well as reduce the space and material required.

In Sec.~\ref{sec:UniformBeam}, the AMM is used to design an rf cavity capable of uniformising a beam by transforming its transverse spatial distribution from Gaussian to uniform. Uniform beams find many applications, including in homogeneous irradiation for processing tasks such as sterilisation, wastewater treatment, material processing, and mutation breeding~\cite{UniformIrradiation}; in high-energy electron radiography~\cite{HEER}; in reducing peak energy deposition density during target irradiation~\cite{Yongke}; and in radiotherapy~\cite{DEFT}. A demonstrated method for achieving beam uniformity is the use of nonlinear focusing forces generated by a multipolar magnet composed of octupolar, dodecapolar, and other higher-order components~\cite{UniformBeam}. Here, we demonstrate how the AMM can replicate this functionality by designing an rf cavity that supports a TM$_{\{4,6\}10}$ mode with embedded octupole and dodecapole components.

The paper concludes in Sec.~\ref{sec:Conclusion} with a summary of the findings and a discussion of future studies that could build upon the presented work or extend the application of the AMM into new directions.

\section{Theory}\label{sec:Theory}

\subsection{Framework and assumptions} \label{sec:Assumptions}

\begin{figure}
    \centering
    \includegraphics[width=0.75\linewidth]{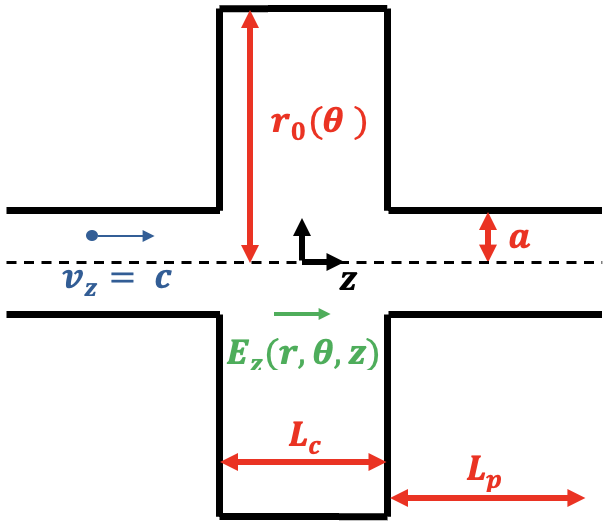}
    \caption{Cross-sectional view of the setup analysed in this paper, showing a particle traversing an azimuthally modulated cavity connected to beam pipes.}
    \label{fig:PillboxSchematic}
\end{figure}

Figure~\ref{fig:PillboxSchematic} shows the setup used for the analysis in this paper, in which a particle of charge $q$ traverses an azimuthally modulated cavity with cross-section $r_0(\theta)$ and cell length $L_c$. The cavity is connected to beam pipes of radii $a$ and lengths $L_p$ and it is assumed to support a harmonic standing-wave TM mode with a longitudinal electric field given by $E_z(\vec{r},t)=E_z(r,\theta,z)\,e^{i(\omega_lt+\psi)}$, where $\omega_l$ is the resonant angular frequency of the mode and $\psi$ is its phase.

To simplify the theoretical analysis, we make the following assumptions:
\begin{itemize}
    \item Rigid particles (or thin cavities): The transverse position $(r, \theta)$ of a given particle is assumed to remain constant as it traverses the cavity. 
    \item Ultra-relativistic and parallel particle: The particles are assumed to have velocity $\vec{v}=c\hat{z}$.
    \item Long beam pipes and evanescent field: The beam pipes are assumed to be sufficiently long that all electromagnetic fields are zero for $|z|>L_p+L_c/2$.
    \item Neglect of fringe fields: Any fringe fields, such as those at the sharp edge between the cavity and beam pipes, are ignored. 
\end{itemize}
Additionally, we locate the centre of the cavity at $z=0$, set the rf phase to be $\psi$ at $t=0$, and set the longitudinal coordinate of the particles to be $z=0$ at $t=0$, such that $t = z/c$.

Under these assumptions, the change in longitudinal momentum can be calculated from the Lorentz force as
\begin{multline}\label{eq:Pz_Change}
    \Delta p_z(r,\theta) = q\int^{L}_{-L} E_z(r,\theta,z,t)dt \\= \frac{q}{c}\int^{\infty}_{-\infty} E_z(r,\theta,z)e^{i(k_lz+\psi)}dz,
\end{multline} 
where $k_l=\omega_l/c$ is the wavenumber of the mode and $c$ is the speed of light. The change in transverse momentum can also be calculated using the Panofsky-Wenzel theorem~\cite{PanofskyWenzel} as
\begin{multline}\label{eq:Pperp_Change}
    \Delta \vec{p}_\perp(r,\theta) 
    = -i\frac{q}{\omega_l}\int_{-L}^{L} \nabla_\perp E_z(\vec{r},t) dz 
    \\= -i\frac{q}{\omega_l}\nabla_\perp\int^{\infty}_{-\infty} E_z(r,\theta,z)e^{i(k_lz+\psi)}dz
    \\ = -i\frac{c}{\omega_l}\nabla_\perp p_z(r,\theta).
\end{multline}

Equations~\ref{eq:Pz_Change}-\ref{eq:Pperp_Change} show that the changes of momentum of particles traversing the cavity are determined solely by the longitudinal component of the electric field. Thus, by tailoring the longitudinal electric field of the cavity, we can directly control both the longitudinal and transverse momentum changes of the particles.

\subsection{Multipolar form of the longitudinal electric field} 

The mathematical form of the electromagnetic field of a TM$_{mnp}$ mode in a pillbox cavity (a cavity with constant cross-section $r_0(\theta)=r_0$ and no beam pipes) is derived in many accelerator textbooks~\cite{Wangler,Lee}. The multipolar form of the longitudinal electric field in such a mode can be written as
\begin{equation}\label{eq:PillboxBasis}
    E_z(r,\theta,z) = \cos{(k_pz)}\tilde{g}_{m} J_{m}\left(\kappa_pr\right)  \cos{(m\theta-\phi_{m})},
\end{equation}
where $\tilde{g}_{m}$ is the magnitude of the multipole component of the mode of order $m$ and $\phi_{m}$ is its orientation ($\phi_m$~=~0 is normal and $\phi_m$~=~$\pi/2$ skew). Here, $J_{m}$ denotes the Bessel function of the first kind of order $m$, $k_p=2p\pi/L_c$, $\kappa_pr_0 = j_{mn}$, where $j_{mn}$ is the $n^\text{th}$ zero of the Bessel function. The integers $n$ and $p$ denote the radial and longitudinal orders of the mode, respectively. 

References~\cite{WroeAMM,WroeThesis} show that Eq.~\ref{eq:PillboxBasis} can be extended to 
\begin{equation}\label{eq:AziBasis}
    E_z(r,\theta,z) = \sum_{\{m\}}\cos{(k_pz)}\tilde{g}_{m} J_{m}\left(\kappa_pr\right)  \cos{(m\theta-\phi_{m})},
\end{equation}
which describes the multipolar form of TM$_{\{m\}\eta p}$ modes composed of a set of multipoles $\{m\}$. Such modes are supported by azimuthally modulated cavities, and the azimuthally modulated method (AMM) provides a means of calculating cavity shapes that support a desired mode. To do so, the boundary condition
\begin{equation} \label{eq:AMMBC}
    0 = \sum_{\{m\}}\cos{(k_pz)}\tilde{g}_mJ_m\left(k_lr_0^{(\eta)}(\theta)\right)\cos{(m\theta-\phi_m)},
\end{equation}
is numerically solved to determine the azimuthally modulated cross-sections $r_0^{(\eta)}(\theta)$ that support specific TM$_{\{m\}\eta p}$ mode. Here, $\eta$ denotes the radial order of the mode and takes a more general definition as the minimum number of radial poles found across all values of $\theta$.

As an example, the TM$_{\{0,1\}\eta,0}$ modes with a monopolar component $g_0~=$~\qty{1}{MV/m} and normal dipole component $g_1~=~$\qty{2}{MV/m} have a longitudinal electric field of the form
\begin{equation}
    E_z(\vec{r}) = 1\times J_0\left(k_lr\right) + 2\times J_1\left(k_lr\right)\cos{\theta}.
\end{equation}
The corresponding set of azimuthally modulated cross-sections $r_0^{(\eta)}(\theta)$ that support such TM$_{\{0,1\}\eta0}$ modes can be found by numerically solving
\begin{equation}
    0 = 1\times J_0\left(k_lr_0^{(\eta)}(\theta)\right) + 2 \times J_1\left(k_lr_0^{(\eta)}(\theta)\right)\cos{\theta},
\end{equation}
for $r_0^{(\eta)}(\theta)$ at each angle $\theta$.

To utilise the AMM in practice, the effect of incorporating beam pipes must be accounted for. To derive this, we begin with a general expression for the longitudinal component of the electric field of a standing-wave mode $l$ in an rf cavity as derived in Ref.~\cite{AbellTransfer}
\begin{equation} \label{eq:Ez}
    E_z(\vec{r}) = \int_{-\infty}^\infty dk\frac{e^{ikz}}{\sqrt{2\pi}}\sum^\infty_{m=0}\tilde{g}_m(k)R_m(\kappa_l r)\cos{(m\theta-\phi_m)}.
\end{equation}
Here, the radial function $R_m(\kappa_l r)$ takes the form of either a regular or a modified Bessel function:
\begin{equation}
    R_m(\kappa_l r) = \begin{cases}
    J_m(\kappa_l r), & |k| < |k_l|; \\
    I_m(\kappa_l r), & \text{otherwise}, \\
    \end{cases}
\end{equation}
where $\kappa_l^2 = |k^2-k_l^2|$.

If the longitudinal electric field is completely known on the surface of a cylinder of radius $R$ as $E_z(R,\theta,z)$, then the multipolar strengths $g_m(k)$ describing the field can be explicitly calculated by a Fourier Transform of Eq.~\ref{eq:Ez} as
\begin{equation} \label{eq:g_m}
    \tilde{g}_m(k) = \frac{1}{R_m(\kappa_l R)}\int_{-\infty}^\infty \frac{dz}{\sqrt{2\pi}}e^{-ikz}E_{z,m}(R,z),
\end{equation}
where $E_{z}(R,\theta,z) = \sum^\infty_{m=0}E_{z,m}(R,z)\cos{(m\theta-\phi_m)}$. 

We assume that the longitudinal electric field along the bore of the azimuthally modulated cavity shown in Fig.~\ref{fig:PillboxSchematic} is given by
\begin{multline} \label{eq:E_Bore}
    E_z(a,\theta,z)=\\
    \begin{cases}
        \sum_{\{m\}}\tilde{G}_{m}\cos{(m\theta-\phi_m)}\cos{(k_pz)}, & |z|<L_c/2;\\
        0, & \text{otherwise},
    \end{cases}
\end{multline}
where $\tilde{G}_{m}$ is the magnitude of the electric field multipolar component in the bore gap. This asserts that the longitudinal electric field varies identically to the case without beam pipes. Accordingly, this bore-gap magnitude can be related to the multipolar magnitude in Eq.~\ref{eq:AziBasis} as
\begin{equation} \label{eq:BoreRel}
    \tilde{G}_m = \tilde{g}_m(k=0)J_m(k_la),
\end{equation}
where, for emphasis, $\tilde{g}_m(k=0)$ corresponds to $\tilde{g}_m$ in Eq.~\ref{eq:AziBasis}.

The assumption outlined above allows the multipolar strengths describing the mode throughout the cavity to be calculated by inserting Eq.~\ref{eq:E_Bore} into Eq.~\ref{eq:g_m}. This gives 
\begin{multline} \label{eq:g_m_Pill}
    \tilde{g}_m(k) = \frac{\tilde{G}_m}{R_m(\kappa_l a)}\int_{-L_c/2}^{L_c/2} \frac{dz}{\sqrt{2\pi}}e^{-ikz} \\
    =
    \frac{L_c\tilde{G}_m}{\sqrt{2\pi}R_m(\kappa_l a)}\frac{\sin{\left(kL_c/2\right)}}{kL_c/2},
\end{multline}
where, for simplicity, we have set $\phi_m$~=~0 and $p=0$, noting that the general case is a straightforward extension obtained by restoring the omitted terms. Substituting Eq.~\ref{eq:g_m_Pill} into Eq.~\ref{eq:Ez} then gives the longitudinal electric field at any point in the cavity as
\begin{multline} \label{eq:Ez_Azi}
    E_z(\vec{r}) = \sum_{\{m\}}\bigg(\frac{L_c\tilde{G}_m}{\sqrt{2\pi}}\cos{(m\theta)} \times \\ 
    \int_{-\infty}^\infty \frac{dk}{\sqrt{2\pi}}e^{ikz}\frac{\sin{\left(kL_c/2\right)}}{kL_c/2}\frac{R_m(\kappa_l r)}{R_m(\kappa_l a)}\bigg). 
\end{multline}
Equation~\ref{eq:Ez_Azi} thus provides the multipolar form of the longitudinal electric field in an azimuthally modulated cavity with beam pipes.


\subsection{Multipolar form of the change in momentum} 

From the multipolar form of the longitudinal electric field, we can derive the corresponding multipolar expression for the change in longitudinal momentum. Substituting Eqn~\ref{eq:Ez_Azi} into Eq.~\ref{eq:Pz_Change} gives
\begin{multline}\label{eq:Pz_Azi}
    \Delta p_z(r,\theta) = \frac{q}{c} \sum_{\{m\}}\frac{L_c\tilde{G}_m}{\sqrt{2\pi}}\cos{(m\theta)}e^{i\psi} \\ 
    \int^{\infty}_{-\infty}\int_{-\infty}^\infty \frac{dkdz}{\sqrt{2\pi}}e^{ikz}e^{ik_lz}\frac{\sin{\left(kL_c/2\right)}}{kL_c/2}\frac{R_m(\kappa_l r)}{R_m(\kappa_l a)}.
\end{multline} 
Then, using the identity
\begin{equation}
\int^{\infty}_{-\infty}e^{ikz}e^{ik_lz}dz=2\pi\delta(k+k_l),
\end{equation}
and the limit
\begin{equation}
    \lim_{\kappa_l \to 0}\left(R_m(\kappa_lr)/R_m(\kappa_la)\right)=\left(r/a\right)^m,
\end{equation}
simplifies Eq.~\ref{eq:Pz_Azi} to the concise multipolar form
\begin{equation}\label{eq:Pz_Azi_Final}
    \Delta p_z(r,\theta) = \frac{q}{c}L_c\frac{\sin{\left(k_lL_c/2\right)}}{k_lL_c/2}\cos{\psi} \sum_{\{m\}}\tilde{G}_m 
    \left(\frac{r}{a}\right)^m\cos{(m\theta)}.
\end{equation} 

Defining the on-axis voltage as
\begin{equation}\label{eq:OnAxisV}
    V_{z,0}~=~\int^{\infty}_{-\infty} E_z(0,0,z)e^{ik_lz}dz=L_c\frac{\sin{\left(k_lL_c/2\right)}}{k_lL_c/2}\tilde{G}_0,
\end{equation}
we note that Eq.~\ref{eq:Pz_Azi_Final} can also be written as
\begin{equation}\label{eq:Pz_Azi_Final2}
    \Delta p_z(r,\theta) = \frac{q}{c}V_{z,0}\cos{\psi} \sum_{\{m\}}\frac{\tilde{G}_m}{\tilde{G}_0} 
    \left(\frac{r}{a}\right)^m\cos{(m\theta)}.
\end{equation}

Finally, we can utilise the Panofsky-Wenzel theorem in Eq.~\ref{eq:Pperp_Change} to obtain the corresponding change in transverse momentum 
\begin{widetext}
\begin{equation}\label{eq:Pperp_Azi_Final}
    \Delta \vec{p}_\perp(r,\theta) = \\\begin{cases} 
        0, & m=0;\\
        \dfrac{q}{\omega_l}\dfrac{L_c}{a}
    \dfrac{\sin{\left(k_lL_c/2\right)}}{k_lL_c/2}\sin{\psi}   \sum\limits_{\{m\}}m\tilde{G}_m\left(\dfrac{r}{a}\right)^{m-1}\begin{pmatrix}\cos{m\theta} \\ -\sin{m\theta}\end{pmatrix}_{(r,~\theta)}, & m\geq1.
    \end{cases}
\end{equation}
Furthermore, by substituting Eqn.~\ref{eq:OnAxisV} and performing a change of basis, Eqn.~\ref{eq:Pperp_Azi_Final} can be expressed  concisely in Cartesian co-ordinates as
\begin{equation}\label{eq:Pperp_Azi_Final_Cart}
    \Delta \vec{p}_\perp(r,\theta) = \\\begin{cases} 
        0, & m=0;\\ \dfrac{q}{\omega_l}\dfrac{V_{z,0}}{a}
    \sin{\psi}  \sum\limits_{\{m\}}m\dfrac{\tilde{G}_m}{\tilde{G}_0}\left(\dfrac{r}{a}\right)^{m-1}\begin{pmatrix}\cos{(m-1)\theta} \\ -\sin{(m-1)\theta}\end{pmatrix}_{(x,~y)}, & m\geq1.\end{cases}
\end{equation} 
\end{widetext}

Equations~\ref{eq:Pz_Azi_Final2} and \ref{eq:Pperp_Azi_Final_Cart} are multipolar expressions for the changes in longitudinal and transverse momentum that provide the desired extension to the AMM framework. The momentum changes are related to the longitudinal electric field via the coefficients $\tilde{G}_m$, which in turn can be related to the required azimuthally cavity cross-section $r_0^{(\eta)}(\theta)$ through Eq.~\ref{eq:BoreRel} and Eq.~\ref{eq:AMMBC}.  

Equations.~\ref{eq:Pz_Azi_Final}-\ref{eq:Pperp_Azi_Final_Cart}  explicitly show that the radial dependence of the momentum change is polynomial, rather than Bessel. This is consistent with previous results for the longitudinal and transverse wakefield potentials in circularly symmetric structures~\cite{Wakefield1,Wakefield2,Wakefield3}.

\section{Example Azimuthally Modulated Cavity} \label{sec:Analysis}

\begin{figure}[t]
\subfloat[][Isometric view.\label{subfig:AziCav3D}]{\includegraphics[width=0.24\textwidth]{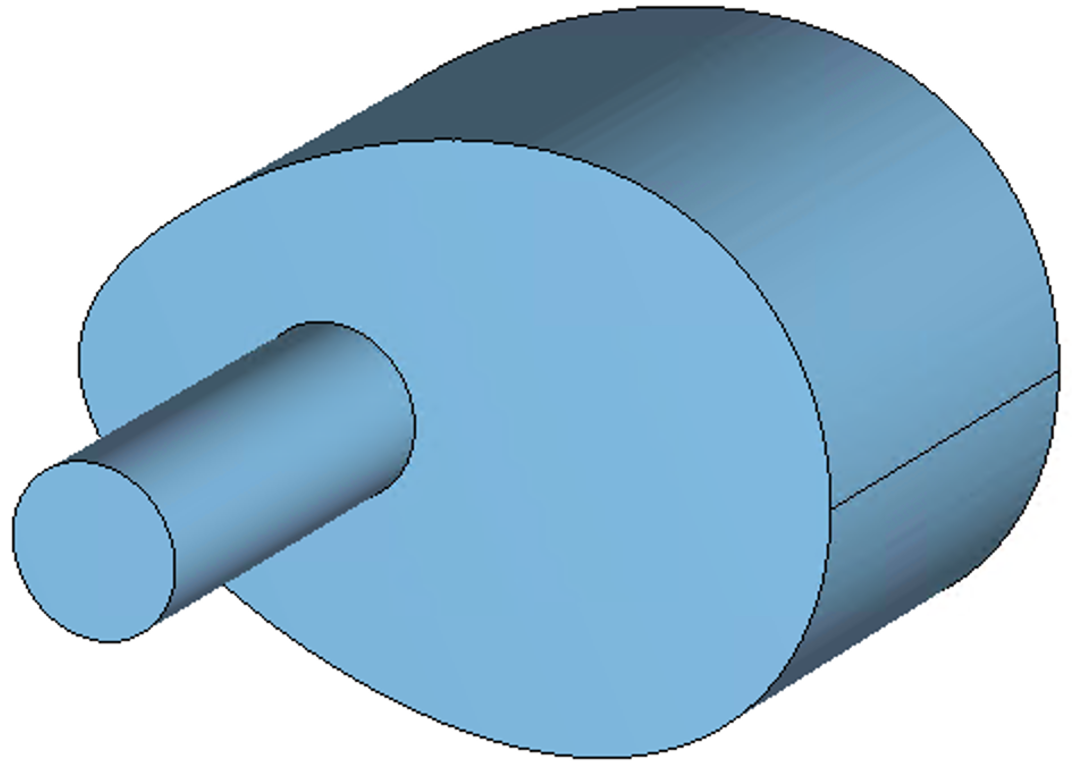}}
\subfloat[][Face-on view.\label{subfig:AziCavX}]{\includegraphics[width=0.24\textwidth]{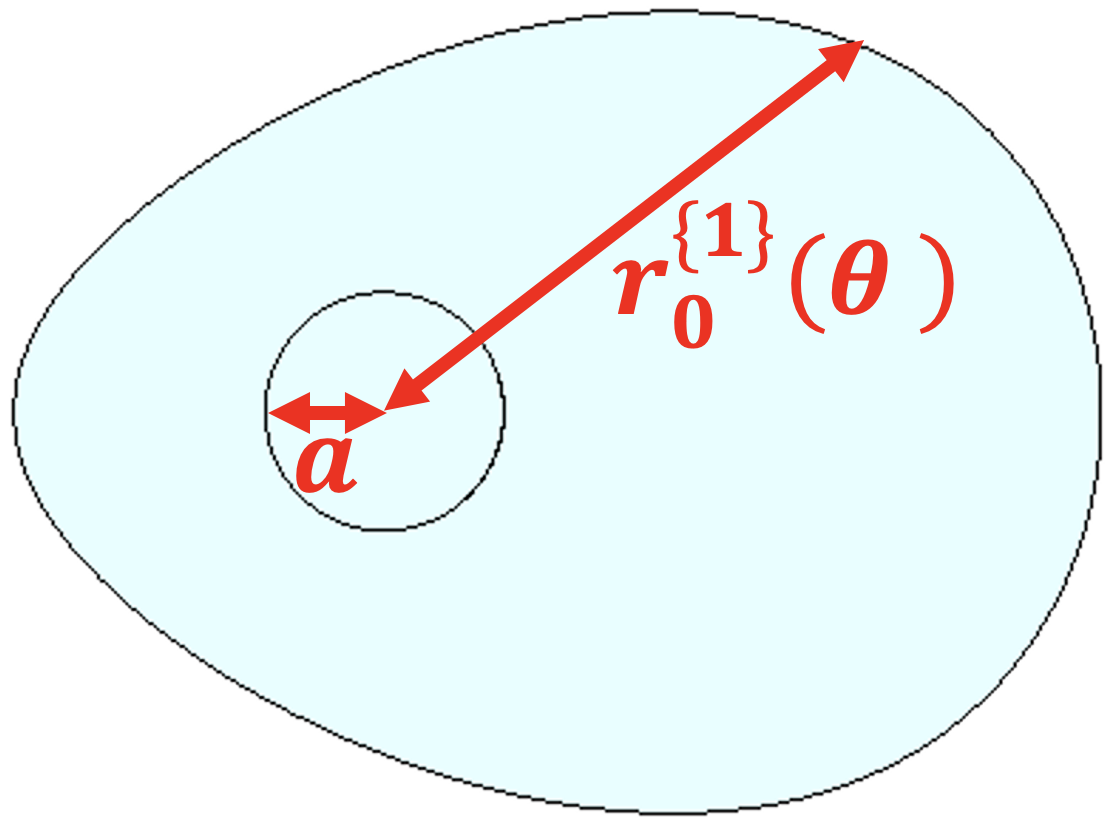}}\\
\subfloat[][Cutting plane view.\label{subfig:AziCavCut}]{\includegraphics[width=0.21\textwidth]{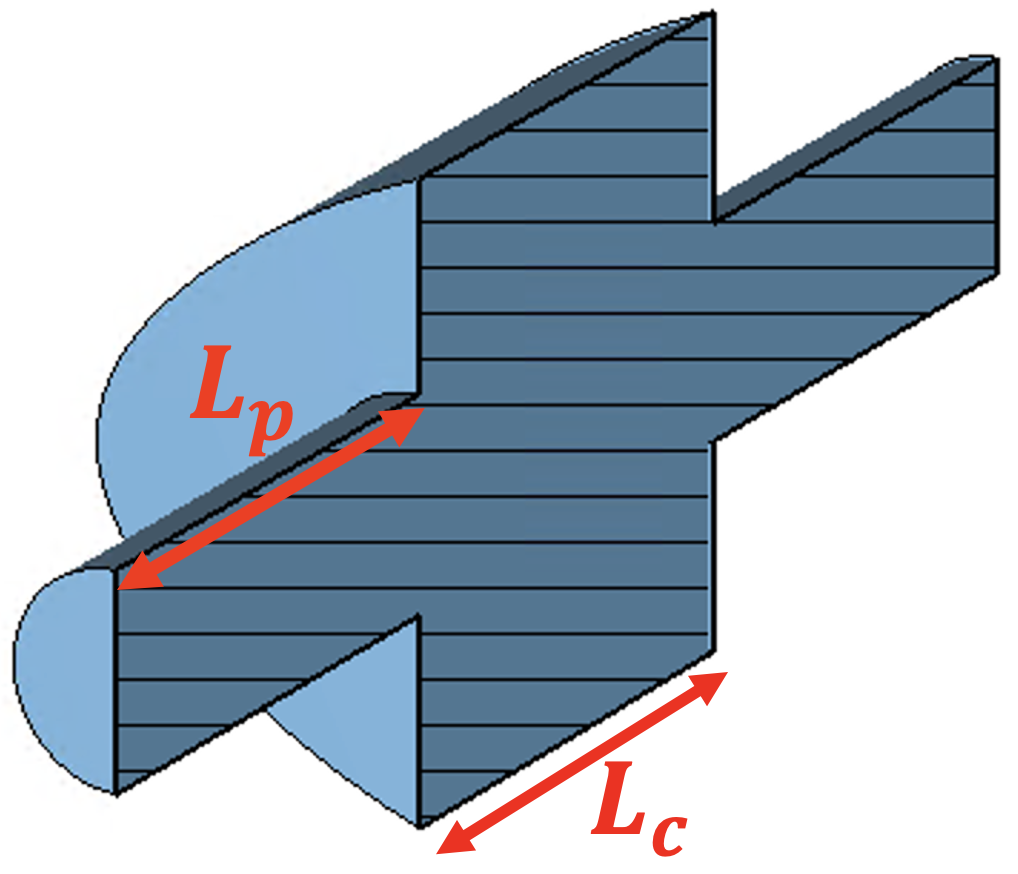}}
\subfloat[][Mode.\label{subfig:AziFieldCav}]{\includegraphics[width=0.28\textwidth]{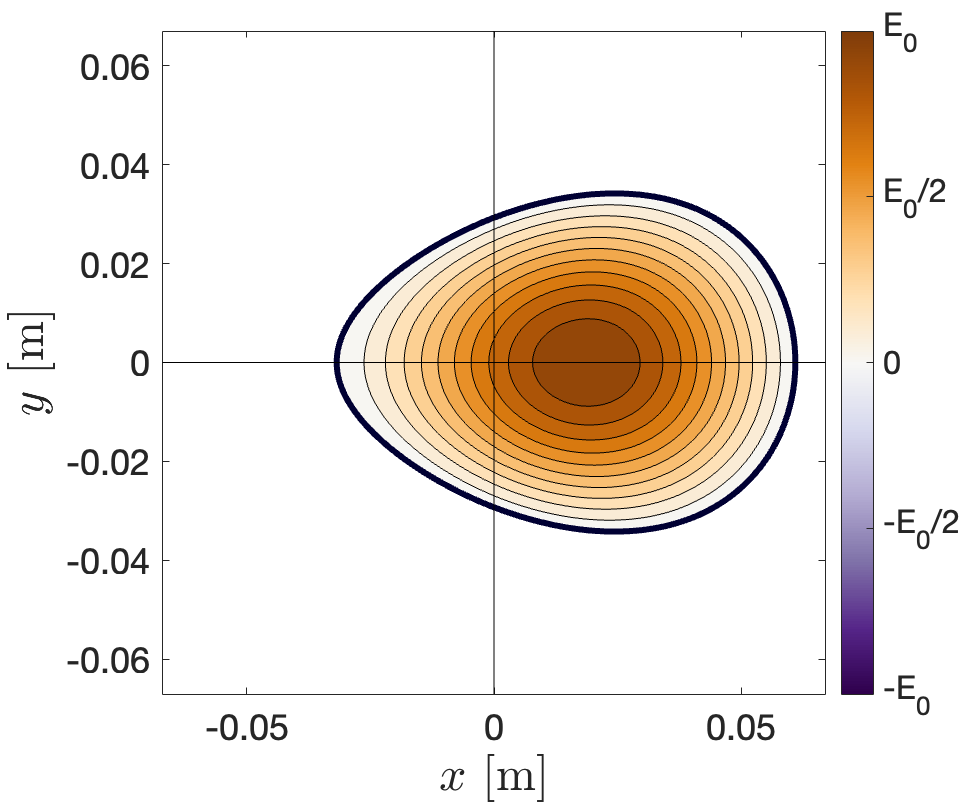}}
\caption{Isometric (a), face-on (b), and cutting plane (c) views of the azimuthally modulated cavity, with beam pipe radius $a$~=~\qty{10}{mm} and cell length $L_c=c/2f_l\simeq$~\qty{50}{mm}, designed to support a \qty{3}{GHz} TM$_{\{0,1,2\}10}$ mode with $\tilde{g}_1/\tilde{g}_0 = \tilde{g}_2/\tilde{g}_0 = $~1. The corresponding contour plot of $E_z$ at the centre of the cavity is shown in (d).}\label{fig:AziCav}
\end{figure}

To analyse the expressions derived in Sec.~\ref{sec:Theory}, we consider a cavity of the type in Fig.~\ref{fig:PillboxSchematic}, with dimensions $L_c=c/2f_l\simeq$~\qty{50}{mm}, $a$~=~\qty{10}{mm}, and $L_p~=~$\qty{50}{mm} that supports a $f_l=$~\qty{3}{GHz} TM$_{\{0,1,2\}10}$ mode with $\tilde{g}_1/\tilde{g}_0 = \tilde{g}_2/\tilde{g}_0 = $~1. The azimuthally modulated cross-section,  $r_0^{\{1\}}(\theta)$, that supports this mode is obtained by solving Eq.~\ref{eq:AMMBC} with the given parameters. The resulting cavity geometry is shown in Fig.~\ref{fig:AziCav}. To compensate for the slight change in resonant frequency introduced by the beam pipes, the cross-section was uniformly scaled by $\sim$~0.2\% to precisely tune the mode to \qty{3}{GHz}. 

\subsection{Field comparisons}

The TM$_{\{0,1,2\}10}$ mode of the constructed cavity was solved using the Eigenmode Solver in the 3D simulation software \textsc{CST}~\cite{cst_mws}, with $<$~\qty{300000}{} mesh cells. \textsc{CST} normalises the computed electromagnetic fields such that the total stored energy in the mode is \qty{1}{J}. For the modelled cavity, this corresponds to an on-axis voltage of $V_{z,0}$~=~\qty{1.24}{MV} and an average on-axis electric field of~\qty{24.8}{MV/m}.

\begin{figure}[t!]
    \centering
    \includegraphics[width=1\linewidth]{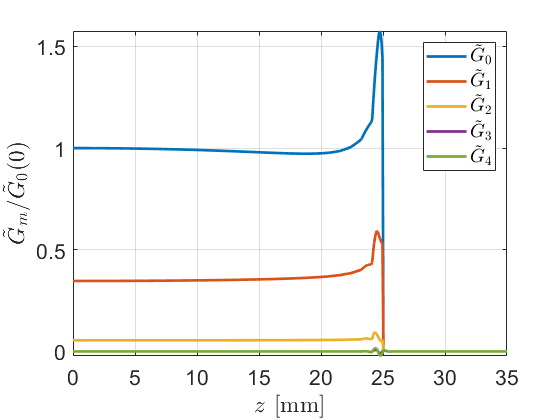}
    \caption{Multipolar contributions to the longitudinal electric field along the bore gap, normalised to the central value of $\tilde{G}_0(0)$.}
    \label{fig:BoreField}
\end{figure}

Figure~\ref{fig:BoreField} shows the multipolar contributions to the longitudinal electric field in the bore gap, normalised to the monopolar contribution at the cavity centre $\tilde{G}_0(0)$. We find that $\tilde{G_m}/\tilde{G}_0$ remains constant throughout the bore gap to within $<~6\%$ for $|z|<$~\qty{20}{mm}, with a notable divergence around the beam pipe intersection at $z=$~\qty{25}{mm}. This deviation is expected due to the mesh coarseness and the sharp corner at the beam pipe intersection.  At the centre of the bore gap, the multipolar contributions are: $\tilde{G}_1/\tilde{G}_0$~=~\qty{0.3466}{} (corresponding to $\tilde{g}_1/\tilde{g}_0$~=~\qty{1.0471}{} using Eq.~\ref{eq:BoreRel}), $\tilde{G}_2/\tilde{G}_0$~=~\qty{0.0547}{} ($\tilde{g}_2/\tilde{g}_0$~=~\qty{1.0338}{}), $|\tilde{G}_3/\tilde{G}_0|~=$~\qty{0.0002}{}, and $|\tilde{G}_4/\tilde{G}_0|~<~$\qty{0.0001}{}. The bracketed values indicate that the AMM design values of $\tilde{g}_1/\tilde{g}_0 = \tilde{g}_2/\tilde{g}_0 = $~1 agree to within $<$~\qty{5}{\%}, supporting the validity of the assumption in Eq.~\ref{eq:E_Bore} that fringe fields can be neglected.  

As an additional comparison, Eq.~\ref{eq:Ez_Azi} was numerically integrated in \textsc{Mathematica}~\cite{Mathematica} using 
the monopole, dipole, and quadrupole components measured at the centre of the bore gap. The integration used 1000 steps in the $dz$-integral up to \qty{100}{mm} and 2000 in the $dk$-integral up to 5000$k_l$. 

\begin{figure}[h!]
\includegraphics[width=0.4\textwidth]{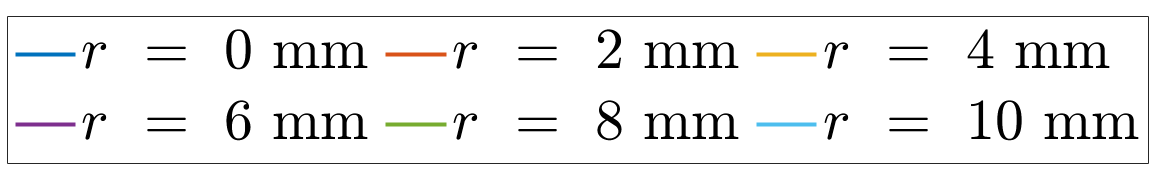}\\
\subfloat[][Static field with $E_z(r,0,z,t=0)$.\label{subfig:EStat}]
{\includegraphics[width=0.45\textwidth]{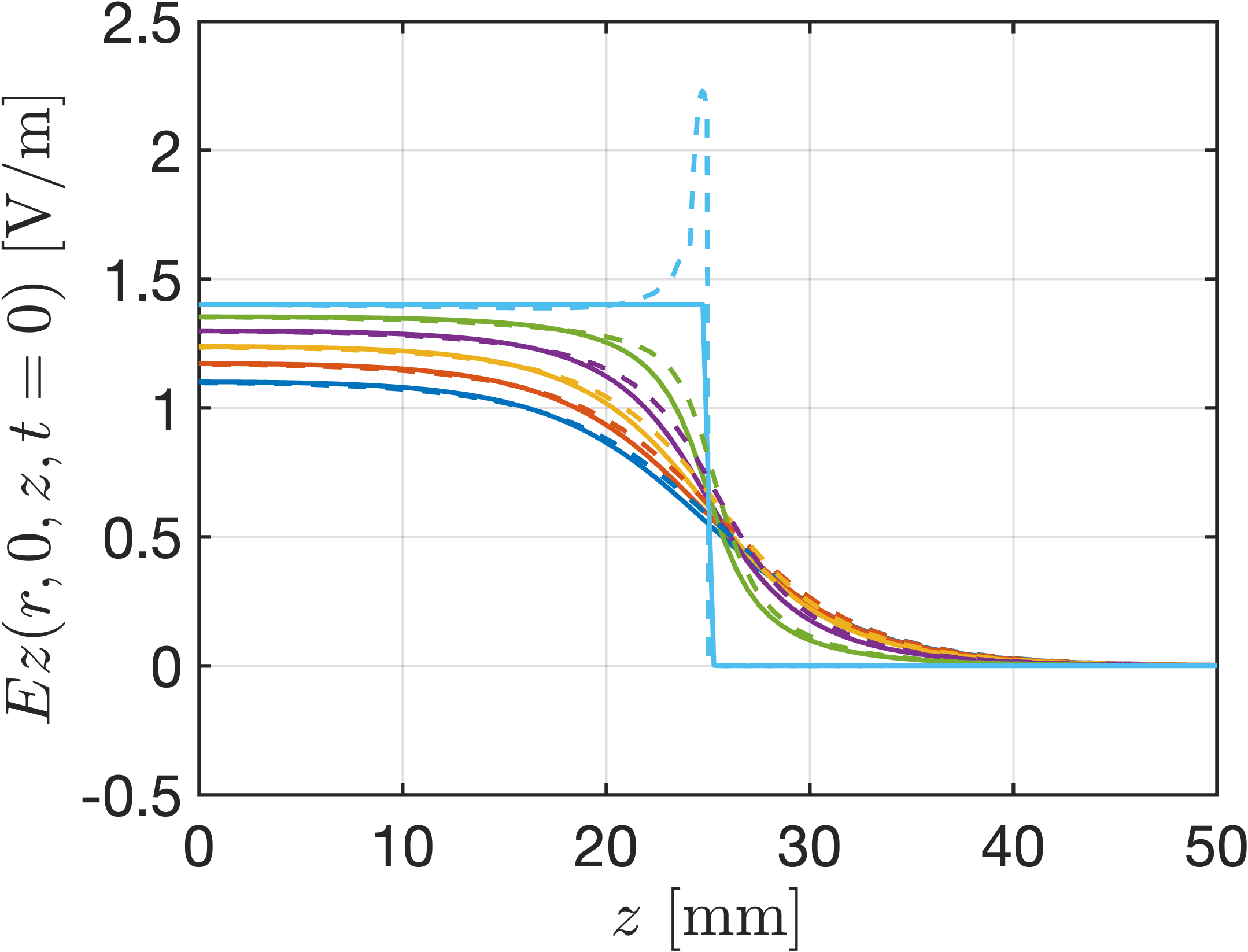}}\\
\subfloat[][Time-varying field with $E_z(r,0,z,t=z/c)$.\label{subfig:ETime}]{\includegraphics[width=0.45\textwidth]{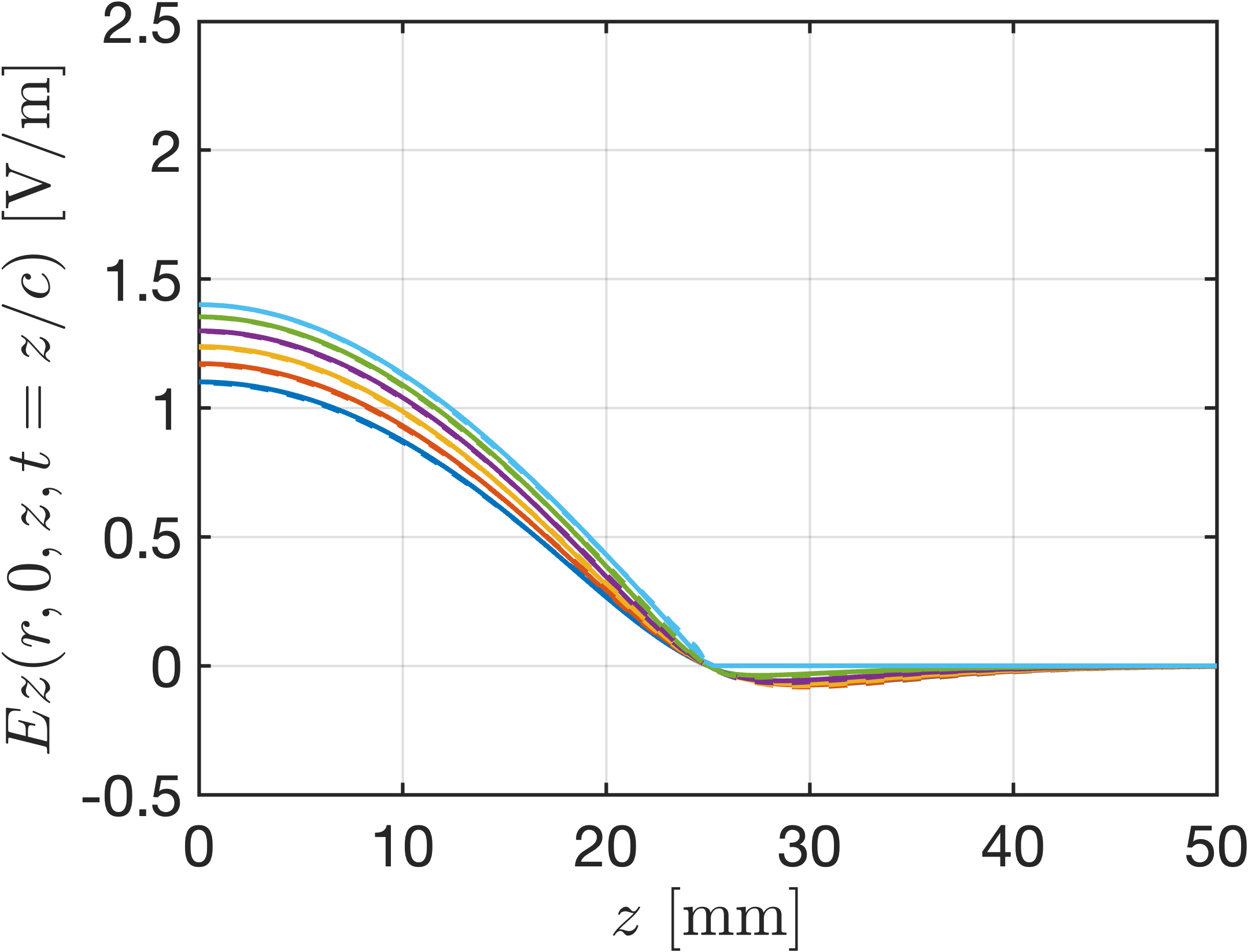}}
\caption{Comparison of the longitudinal electric field along the different radial lines described in the legend for the static (a) and time-varying (b) TM$_{\{{0,1,2}\}10}$ mode, the latter representing the field as seen by an ultra-relativistic particle. Results are computed using \textsc{CST} (dashed lines) and by numerical integration of Eq.~\ref{eq:Ez_Azi} (including time-variation) using \textsc{Mathematica} (solid lines).}\label{fig:MathematicaCST}
\end{figure}

Figure~\ref{fig:MathematicaCST} compares the longitudinal electric field profiles calculated using \textsc{Mathematica} and \textsc{CST} at different radial positions for $\theta=0$. Figure~\ref{subfig:EStat} shows that,  there is a $<$~5~\% difference between the two results for $z<$~\qty{20}{mm} when the time-variation of the field is neglected. Figure~\ref{subfig:ETime} shows that including time-variation reduces this difference and, as a result, the computed $\Delta p_z$ from numerical integration, \textsc{CST} simulation, and Eq.~\ref{eq:Pz_Azi_Final} all agree to within $<1~\%$.

\subsection{Beam dynamics}

To further explore the expressions derived in Section~\ref{sec:Theory}, we performed beam dynamics simulations using \textsc{RF-Track}~\cite{RF-Track}, a versatile tracking code that numerically integrates the particle trajectories and supports both time-based and space-based tracking. For this investigation, the field map solved using \textsc{CST}  was imported into \textsc{RF-Track} with a step-size of \qty{0.25}{mm}. To probe the field map, a bunch of electrons consisting of 6 concentric rings of radii between $r~$=~\qtyrange{0}{8}{mm} was initialised at the longitudinal position $z_i~=~-(L_c/2+L_p)\simeq-$\qty{75}{mm}, with a momentum $\vec{p}_i = p_{i,z}\hat{z}$. The bunch was then tracked through to $z_f~=+(L_c/2+L_p)$. The change in momentum for each particle was calculated as $\Delta p_z = p_{f,z}-p_{i,z}$ with $\psi~=~\ang{0}$ (on-crest) and $\Delta p_x = p_{f,x}$ and $\Delta p_y = p_{f,y}$ with $\psi~=~\ang{90}$.


As a first test, we tracked an ultra-relativistic beam by setting $\vec{p}_i=p_{i,z}\hat{z}$ and $p_{i,z}~=~$\qty{10}{GeV/c}.  The analytical predictions were calculated using the \textsc{RF-Track} measured value of $V_{z,0}$ and multipolar ratios of $\tilde{G}_1/\tilde{G}_0$~=~$0.3546$ ($\tilde{g}_1$/$\tilde{g}_0$~=~1.0713) and $\tilde{G}_2/\tilde{G}_0$~=~$0.0556$ ($\tilde{g}_2$/$\tilde{g}_0$~=~1.0508). In this case, these latter values were obtained by minimising the root mean square deviation (RMSD) between the analytical (Eq.~\ref{eq:Pz_Azi_Final2}) and simulated values (\textsc{RF-Track}) of $\Delta p_z$ as
\begin{equation}
    \text{RMSD}(\Delta p_z) = \sum_{i}\frac{(\Delta p_z^\text{Analytical}(\theta_i)-\Delta p_z^\textsc{RF-Track}(\theta_i))^2}{n},
\end{equation}
evaluated on the cylinder surface of radius $R~=~$\qty{8}{mm}, where $n$ is the number of samples.

\begin{figure*}[p]
\includegraphics[width=0.75\textwidth]{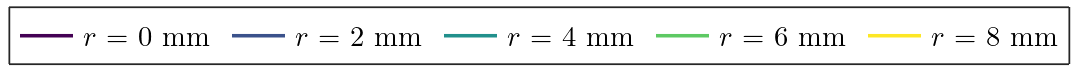}\\
\subfloat[][$p_{i,z}~=~$\qty{10}{GeV/c}, $\psi~=~$\ang{0}.\label{subfig:Pz}]{\includegraphics[width=0.25\textwidth]{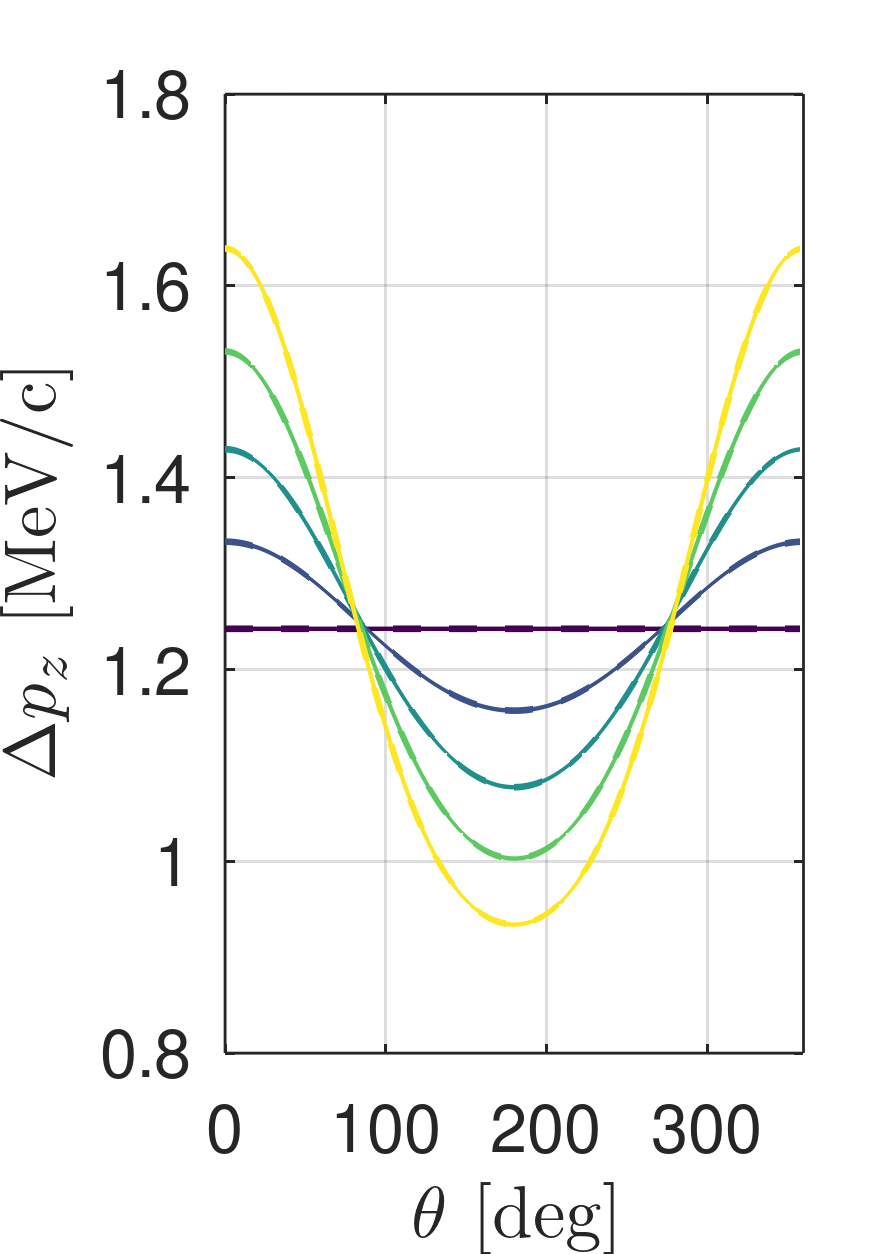}}
\subfloat[][$p_{i,z}~=~$\qty{10}{GeV/c}, $\psi~=~$\ang{90}.\label{subfig:Px}]{\includegraphics[width=0.25\textwidth]{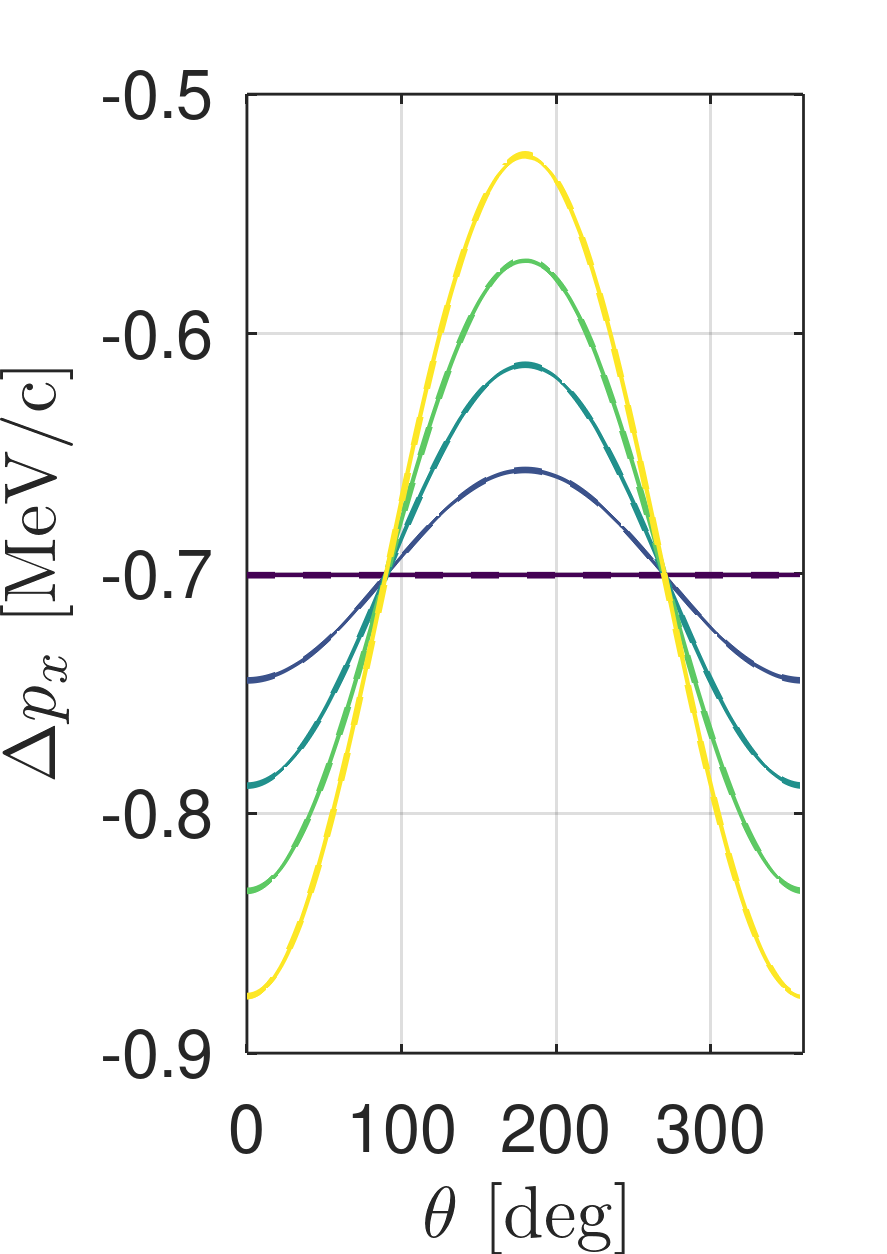}}
\subfloat[][$p_{i,z}~=~$\qty{10}{GeV/c}, $\psi~=~$\ang{90}.\label{subfig:Py}]{\includegraphics[width=0.25\textwidth]{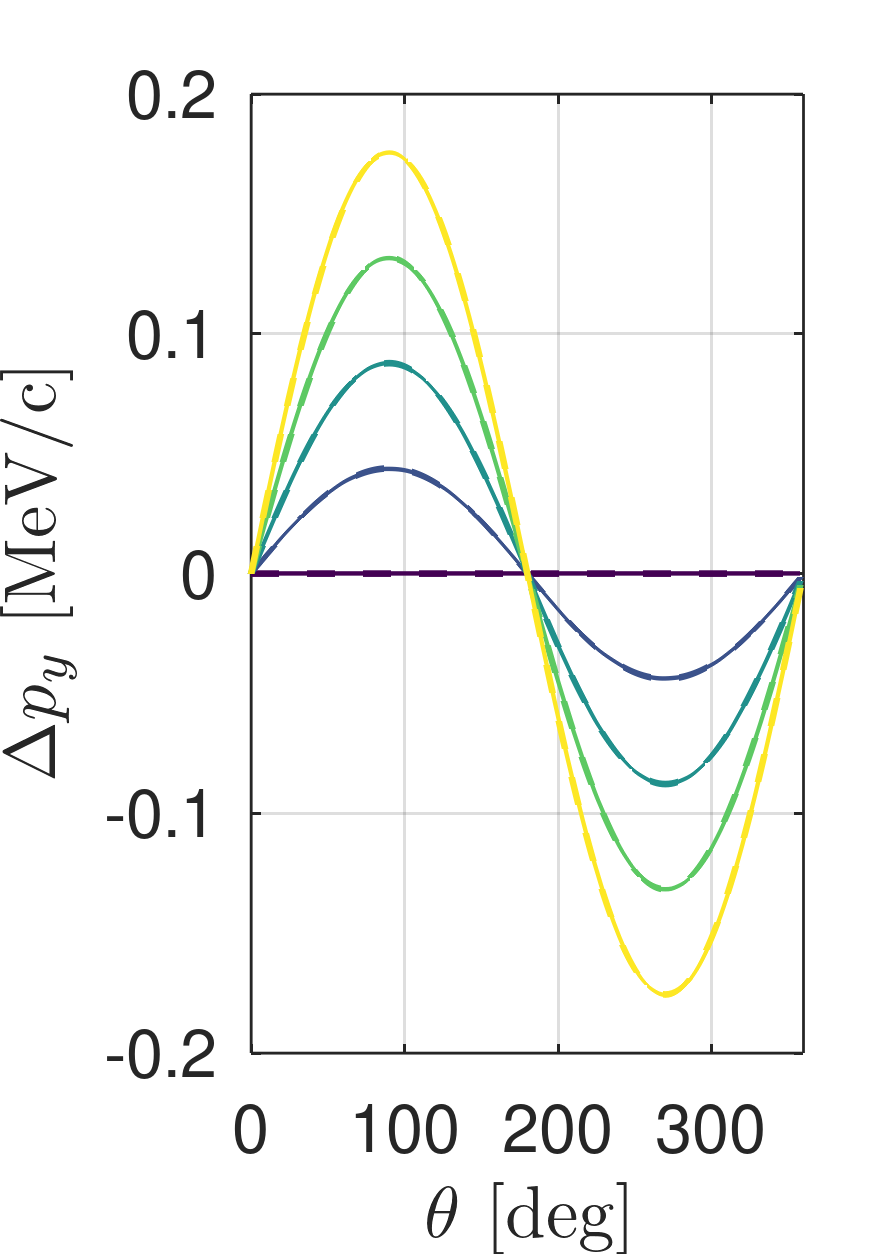}}\\
\subfloat[][$p_{i,z}~=~$\qty{10}{MeV/c}, $\psi~=~$\ang{0}.\label{subfig:Pz_10MeV}]{\includegraphics[width=0.25\textwidth]{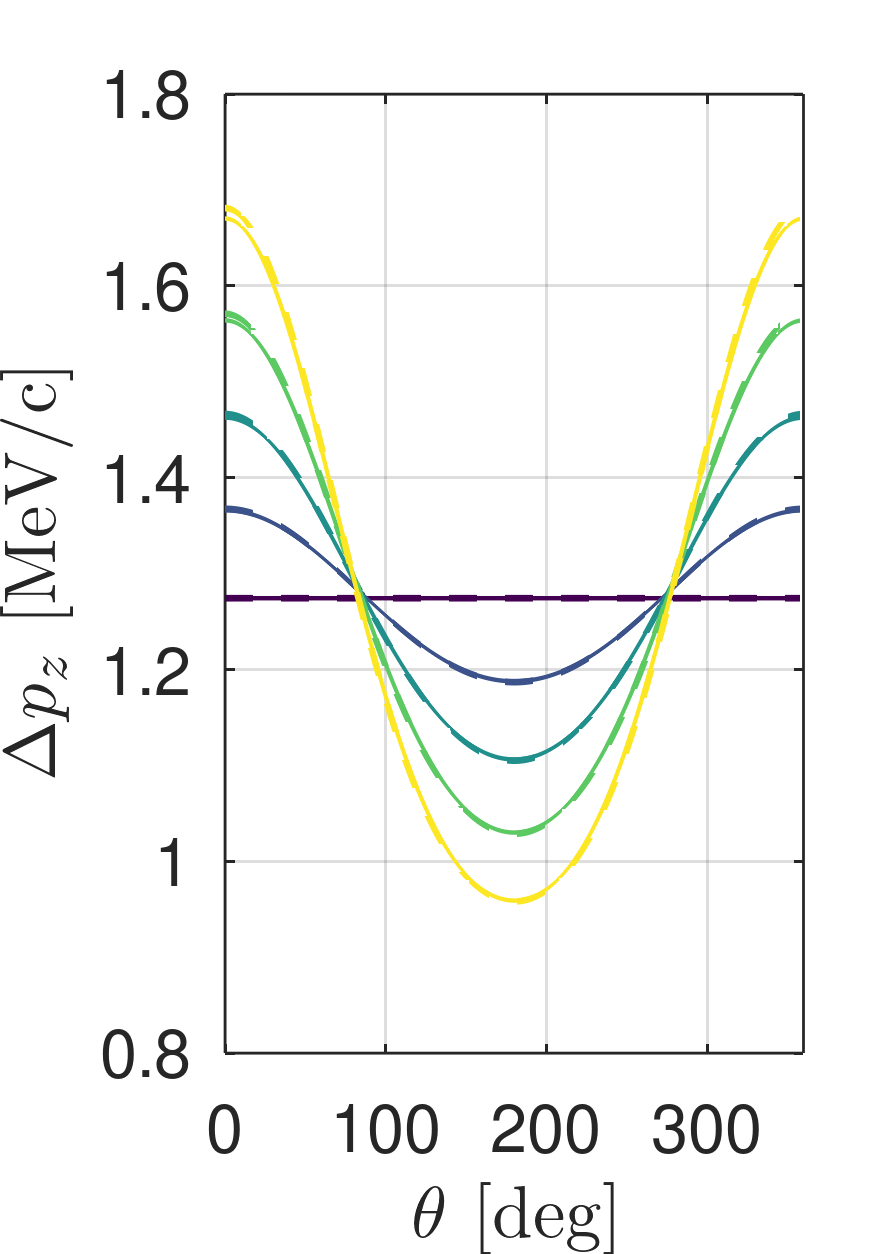}}
\subfloat[][$p_{i,z}~=~$\qty{10}{MeV/c}, $\psi~=~$\ang{90}.\label{subfig:Px_10MeV}]{\includegraphics[width=0.25\textwidth]{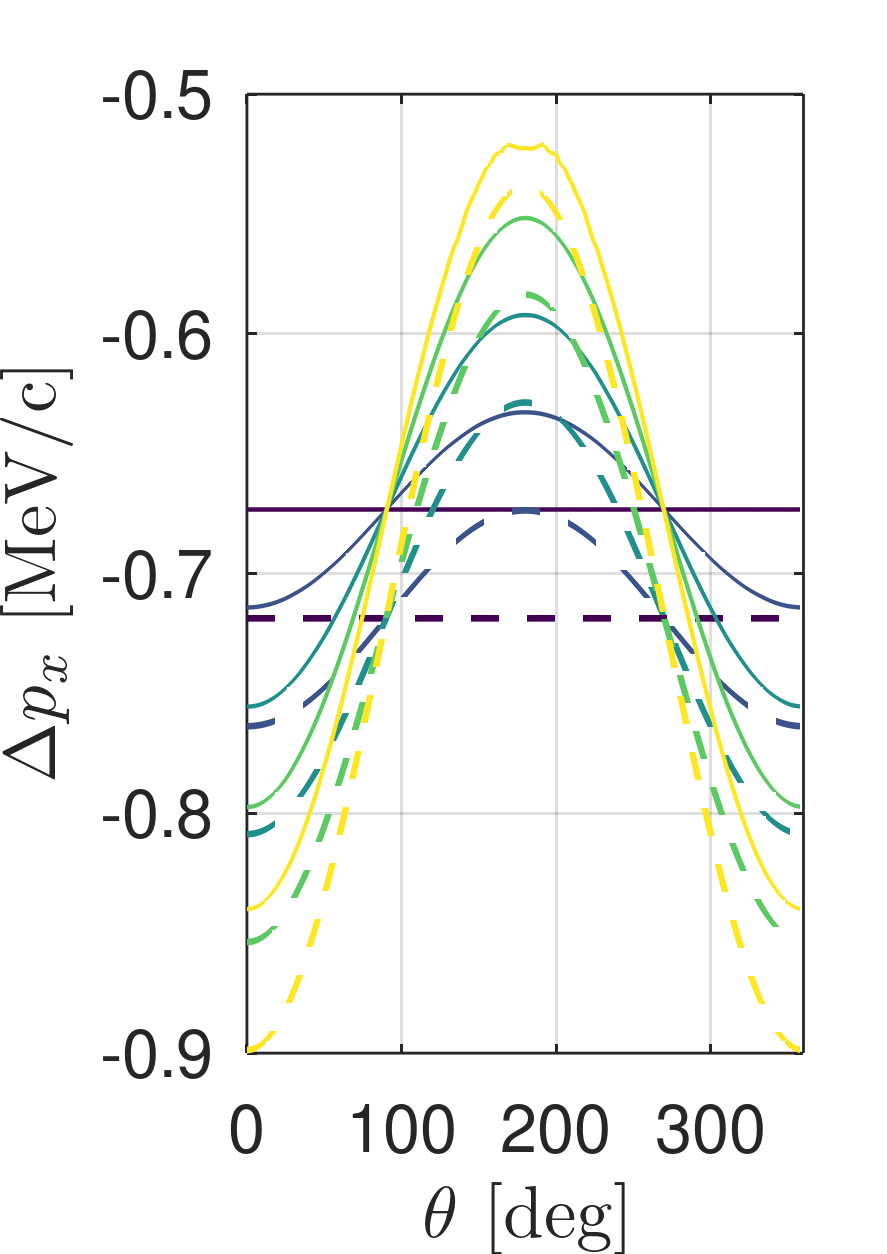}}
\subfloat[][$p_{i,z}~=~$\qty{10}{MeV/c}, $\psi~=~$\ang{90}.\label{subfig:Py_10MeV}]{\includegraphics[width=0.25\textwidth]{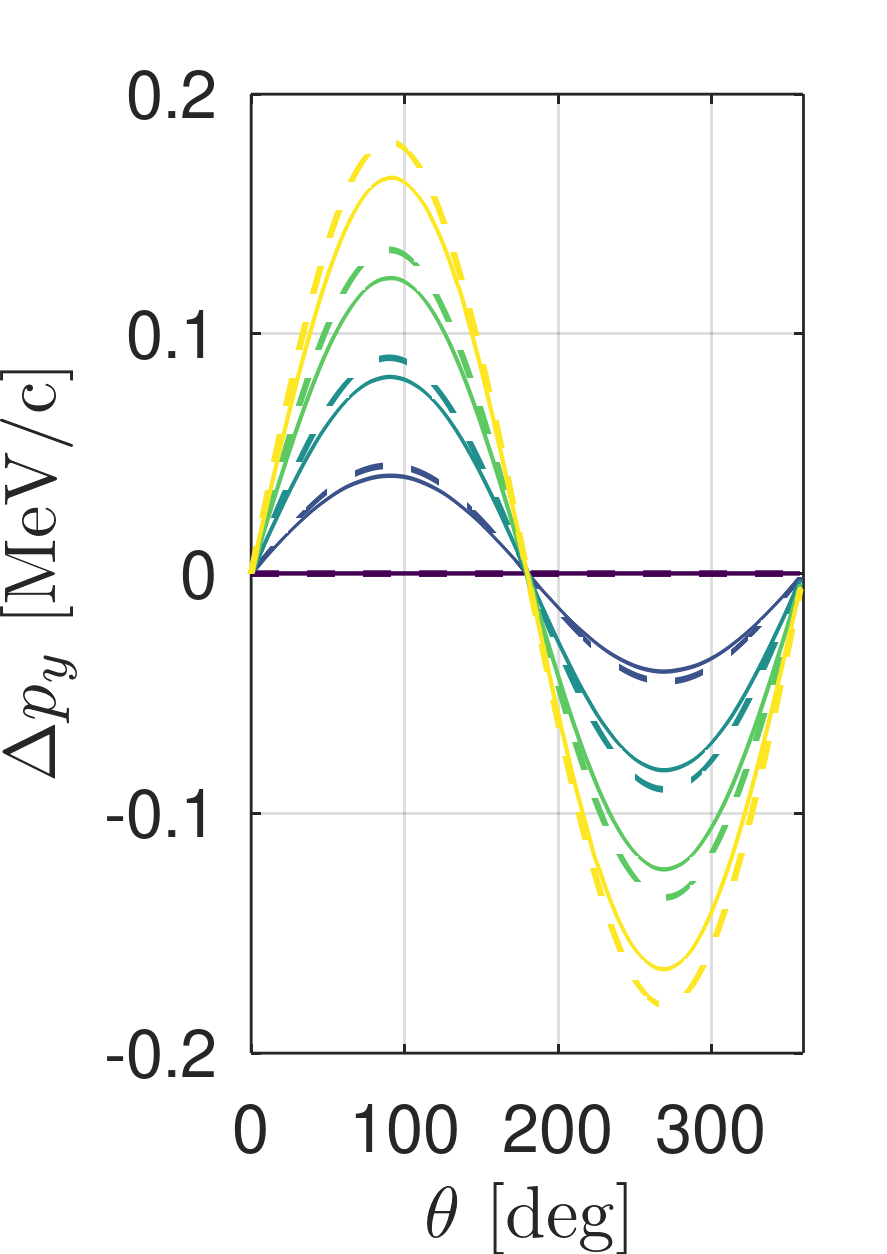}}\\
\subfloat[][$p_{i,z}~=~$\qty{1}{MeV/c}, $\psi~=~$\ang{0}.\label{subfig:Pz_1MeV}]{\includegraphics[width=0.25\textwidth]{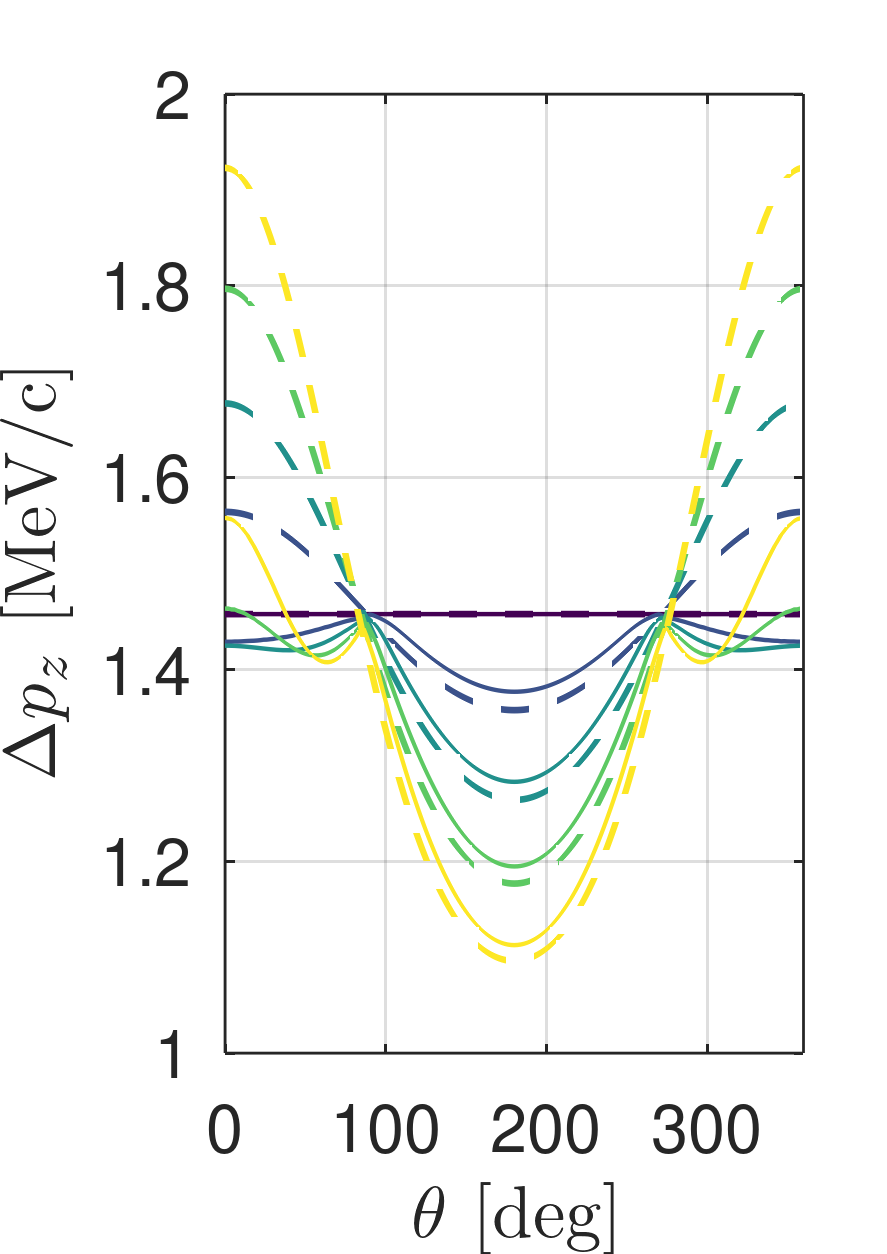}}
\subfloat[][$p_{i,z}~=~$\qty{1}{MeV/c}, $\psi~=~$\ang{90}.\label{subfig:Px_1MeV}]{\includegraphics[width=0.25\textwidth]{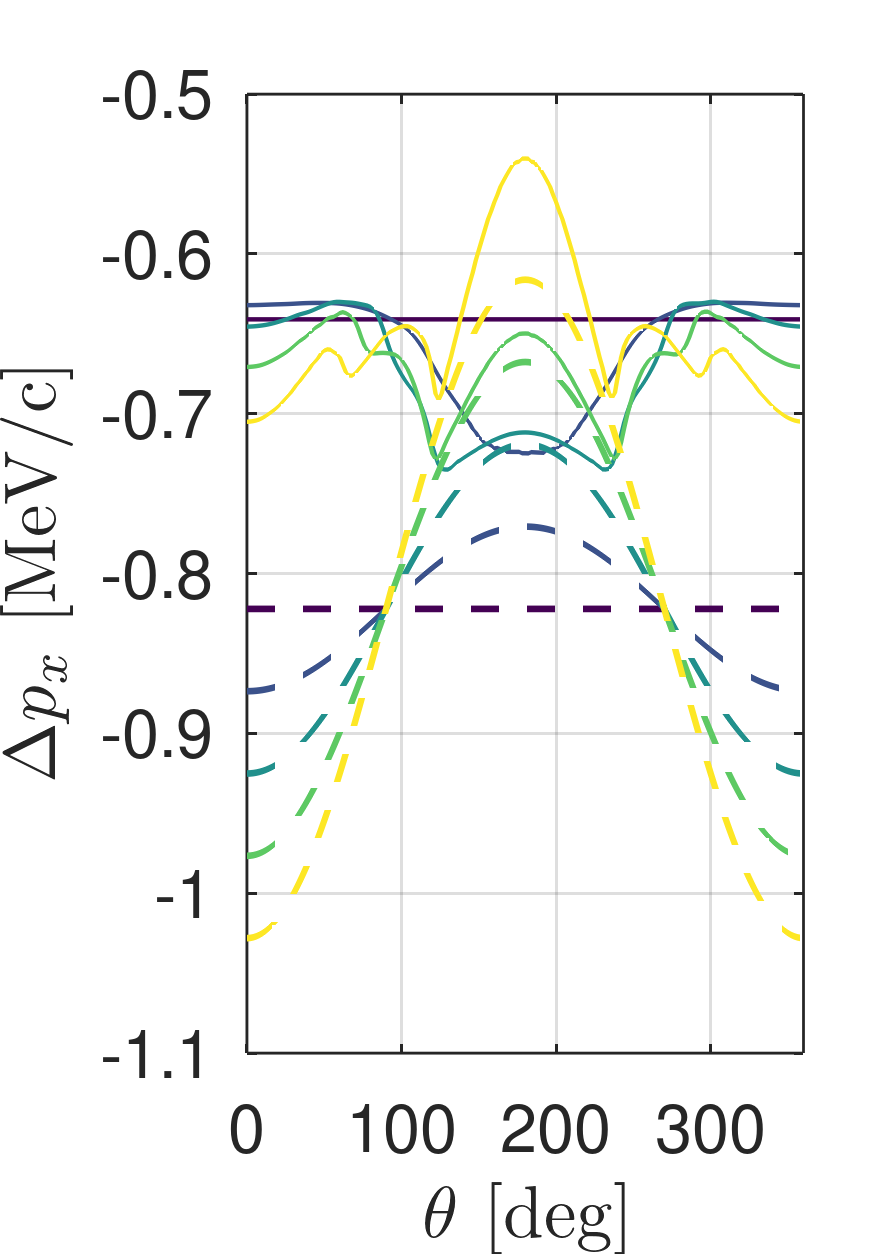}}
\subfloat[][$p_{i,z}~=~$\qty{1}{MeV/c}, $\psi~=~$\ang{90}.\label{subfig:Py_1MeV}]{\includegraphics[width=0.25\textwidth]{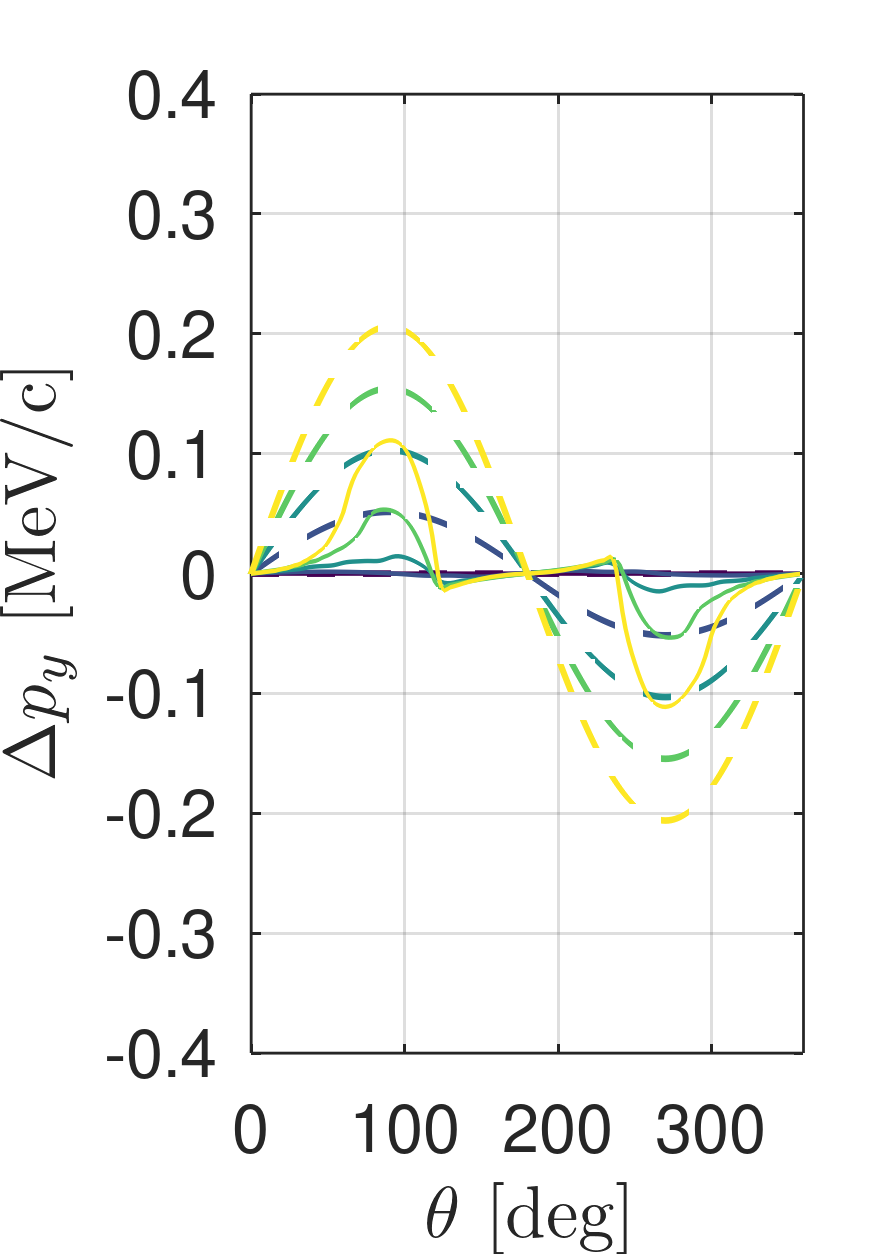}}
\caption{Change in longitudinal, horizontal, and vertical momenta of an electron beam with $\vec{p}_i=p_{i,z}\hat{z}$ for different energies, as computed using \textsc{RF-Track} simulations (solid lines) and calculated from Eqns.~\ref{eq:Pz_Azi_Final2} and \ref{eq:Pperp_Azi_Final_Cart} (dashed lines).}\label{fig:P2_Comparisons}
\end{figure*}

Figures~\ref{subfig:Pz},~\ref{subfig:Px},~and~\ref{subfig:Py} show very good agreement between the analytical formula and \textsc{RF-Track} measurements, with the dashed lines visually coinciding with the solid lines. Quantitatively, the percentage error on the maximum difference between the two results is $<$~\qty{0.7}{\%} for all three components of $\Delta p$.


For the next investigations, we relaxed the assumption of an ultra-relativistic beam. We expect the agreement between analytical formula and \textsc{RF-Track} to degrade as the beam's velocity decreases because:
\begin{itemize}
    \item The change in transverse position of the particle as it traverses the cavity becomes significant. Specifically, the assumption that $\int E_z(r,\theta,z)dt~=~\int E_z(r_i,\theta_i,z)dz$ becomes less  valid.
    \item The change in the longitudinal velocity of the particle as it traverses the cavity becomes significant. Specifically, the assumption that $t~=~z/c$ becomes invalid.
\end{itemize}

First, we compare the change in momentum calculated from the analytical expressions and from \textsc{RF-Track} for an initial momentum of $p_{i,z}~=~$\qty{10}{MeV/c}. In Fig.~\ref{subfig:Pz_10MeV}, we observe good agreement between analytical prediction and simulation for the longitudinal momentum change $\Delta p_z$. In contrast, Figs.~\ref{subfig:Px_10MeV} and \ref{subfig:Py_10MeV} show a clear discrepancy in the transverse components, which arises from the breakdown of assumptions underlying the Panofsky-Wenzel theorem at lower energies. This is most noticeable in Fig.~\ref{subfig:Px_10MeV}, where a discrepancy of approximately \qty{5}{\%} is observed in the constant offset associated with the dipole component.

Upon further lowering the initial momentum to $p_{i,z}~=~$\qty{1}{MeV/c}, Fig.~\ref{subfig:Pz_1MeV} shows  a strong disagreement in $\Delta p_z$, which now arises because the change in longitudinal momentum of the particle is comparable to the initial momentum of the particle. As expected, Figs.~\ref{subfig:Px_1MeV} and \ref{subfig:Py_1MeV}  show very poor agreement for $\Delta p_x$ and $\Delta p_y$ respectively.

\begin{figure*}[tbh!]
\includegraphics[width=0.75\textwidth]{Figures/LegendField.png}\\
\subfloat[][$\psi~=~$\ang{0}.\label{subfig:RMSD_Pz}]{\includegraphics[width=0.25\textwidth]{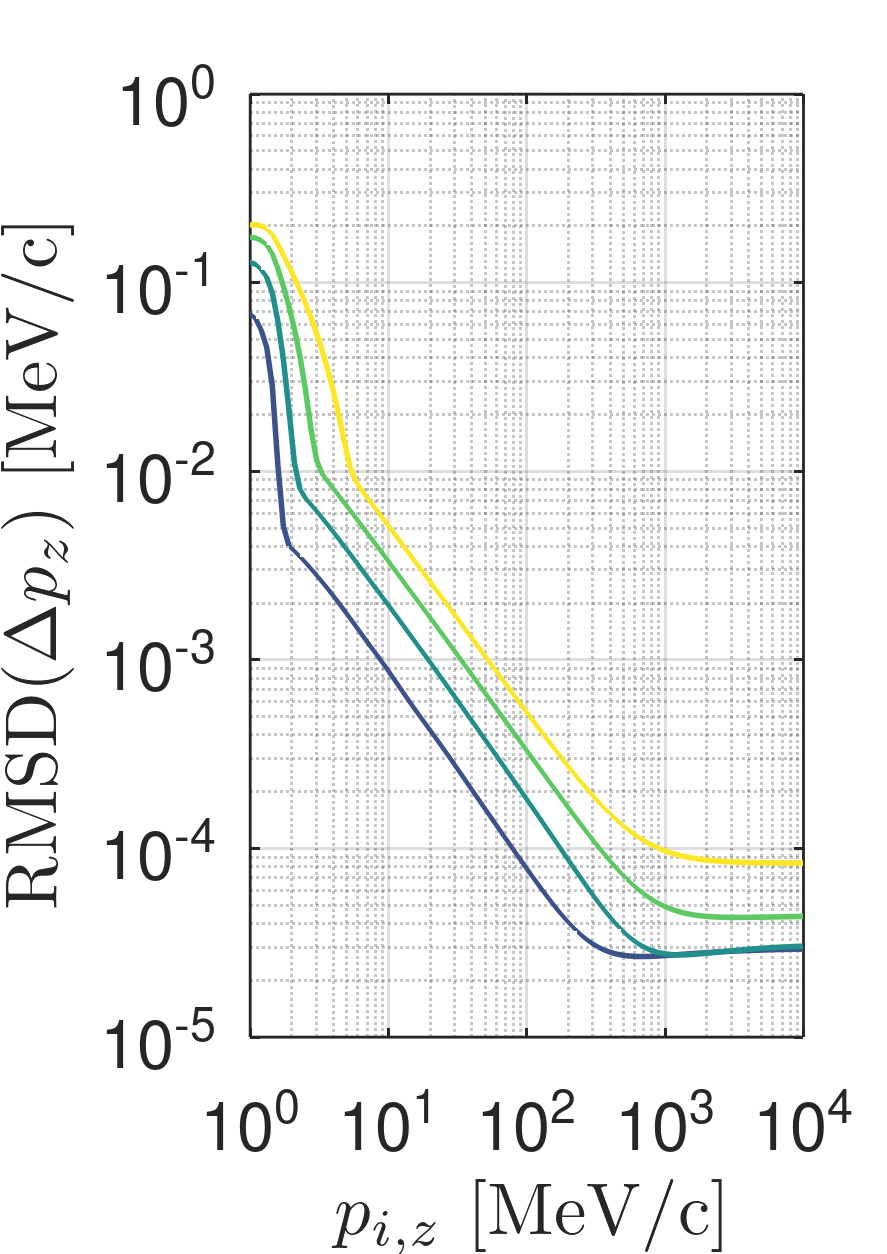}}
\subfloat[][$\psi~=~$\ang{90}.\label{subfig:RMSD_Px}]{\includegraphics[width=0.25\textwidth]{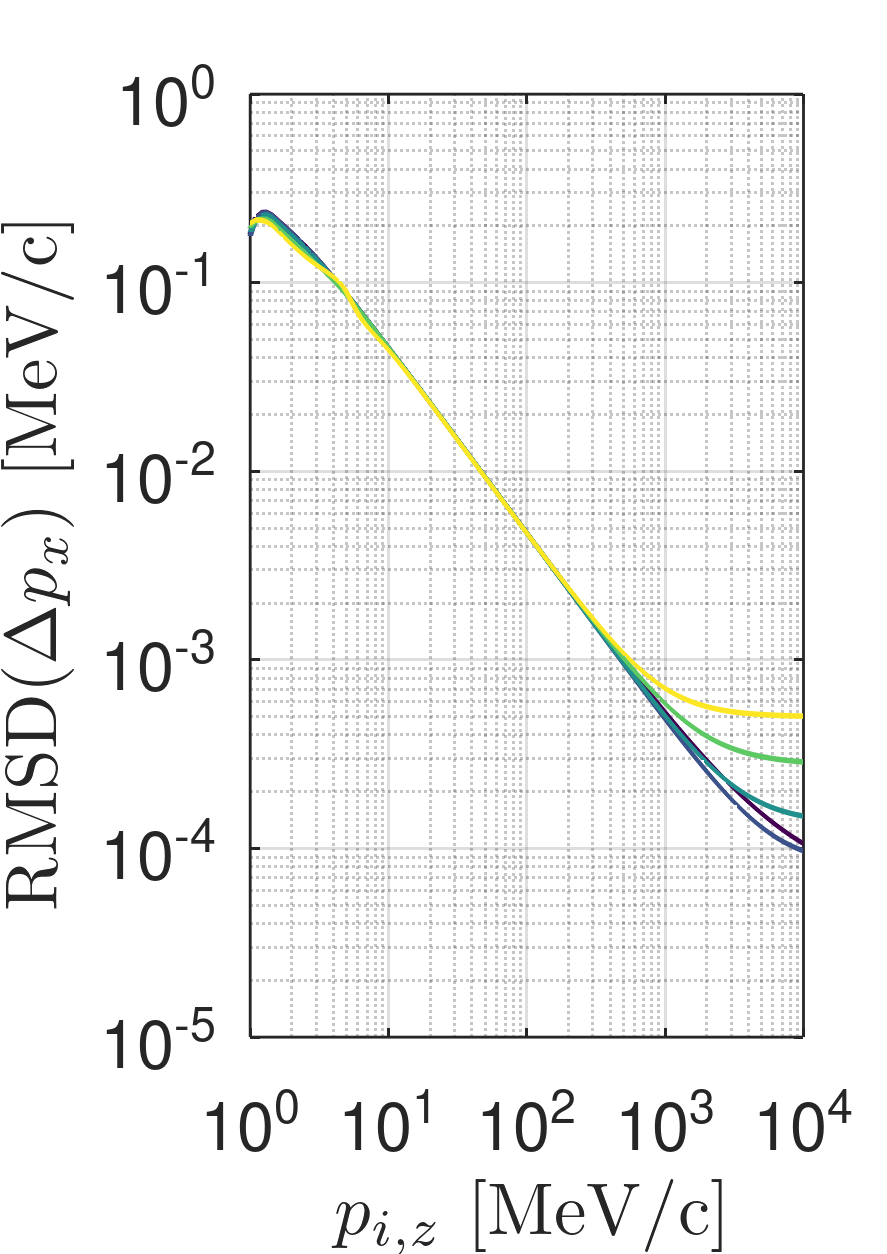}}
\subfloat[][$\psi~=~$\ang{90}.\label{subfig:RMSD_Py}]{\includegraphics[width=0.25\textwidth]{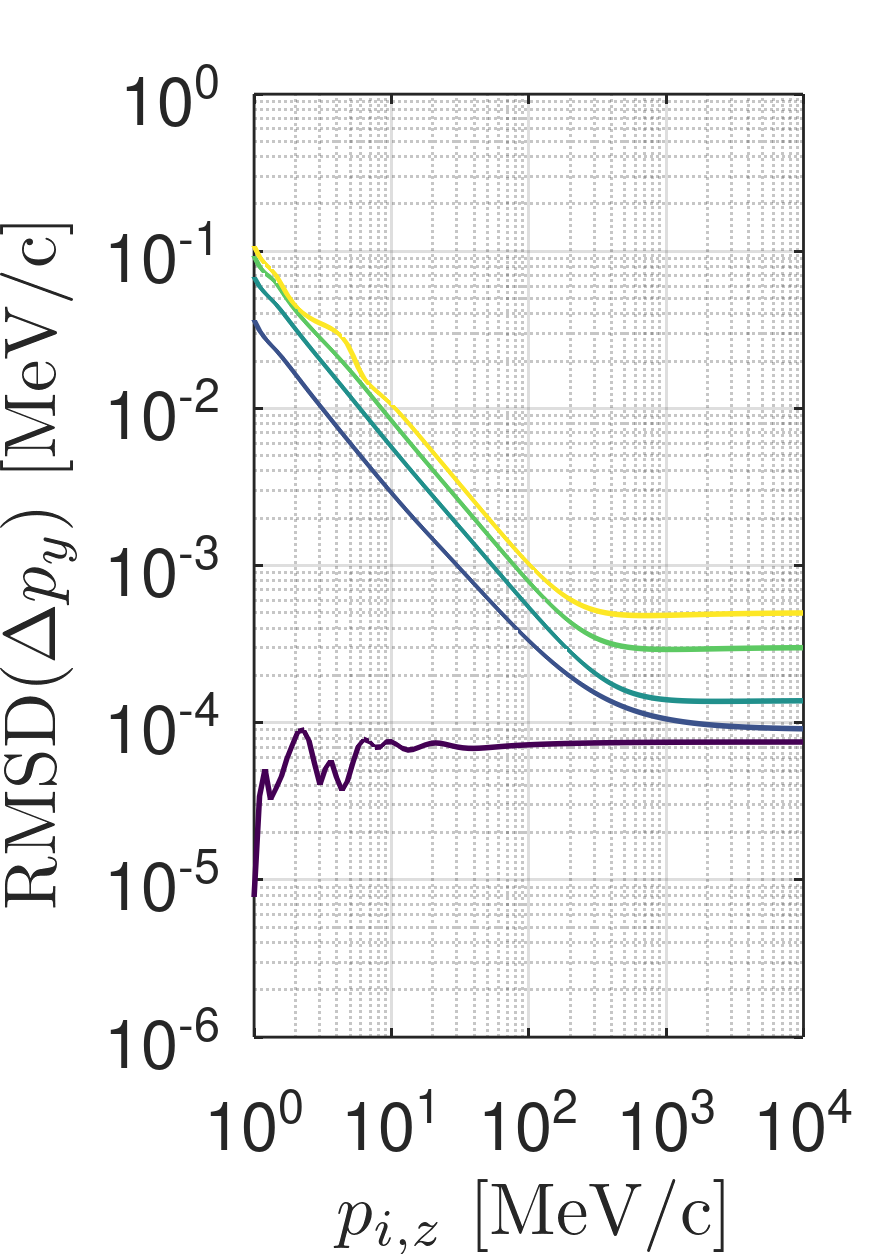}}
\caption{Root mean square deviation (RMSD) between \textsc{RF-Track} simulations and analytical Eqns.~\ref{eq:Pz_Azi_Final2} and \ref{eq:Pperp_Azi_Final_Cart} for the change in longitudinal (a), horizontal (b), and vertical (c) momentum of an electron beam with $\vec{p}_i=p_{i,z}\hat{z}$.}\label{fig:P_RMSD}
\end{figure*}

To visualise the validity of the analytical expressions, Fig.~\ref{fig:P_RMSD} shows the RMSD as a function of initial momentum over the range \qtyrange{1}{10000}{MeV/c}. We observe that $\Delta p_x$ diverges more rapidly than $\Delta p_y$ because the dipole component is orientated in the $x$-direction. 

Nonetheless, we have demonstrated in this section that the derived analytical expressions are in excellent agreement with simulation results at ultra-relativistic energies, and still offer reasonable accuracy at the moderately relativistic energy of \qty{10}{MeV} ($\gamma~\sim$~20).

\section{Design of a Multipole-Free Accelerating Structure}\label{sec:MultipoleFree}

Incorporating ancillaries such as slot-based power couplers and higher-order mode dampers into accelerating rf cavities alters the azimuthal symmetry of the cavity. As a result, unwanted transverse multipolar components are introduced into the TM$_{010}$ mode, transforming it from TM$_{010}$-exact to TM$_{010}$-like. These transverse multipoles can have significant and detrimental effects by introducing transverse deflecting forces~\cite{Lapostolle1970}. 

Multiple mitigation strategies exist to counteract these transverse multipoles. For example, the lower-order components can be negated by raising the order of the azimuthal symmetry of the cavity, such as by including additional slot-based power couplers. Alternatively, the transverse multipoles with a single-slot can be compensated for by offsetting the coupler cavity, or the deflecting effect can be reduced by feeding successive accelerator structures from alternating directions. Here we show how the AMM can be used to completely remove the transverse multipoles from the mode with a single-port coupler incorporated.

\subsection{$N$-port coupler designs}

\begin{figure}[b!]
\subfloat[][1-port coupler.\label{subfig:1_Port}]{\includegraphics[width=0.16\textwidth]{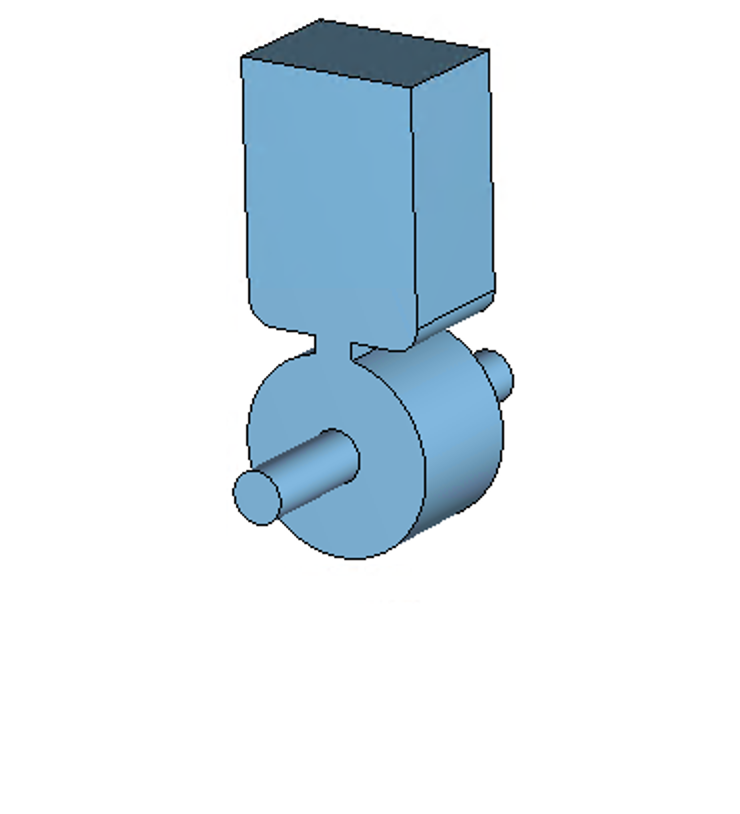}}
\subfloat[][2-port coupler.\label{subfig:2_Port}]{\includegraphics[width=0.16\textwidth]{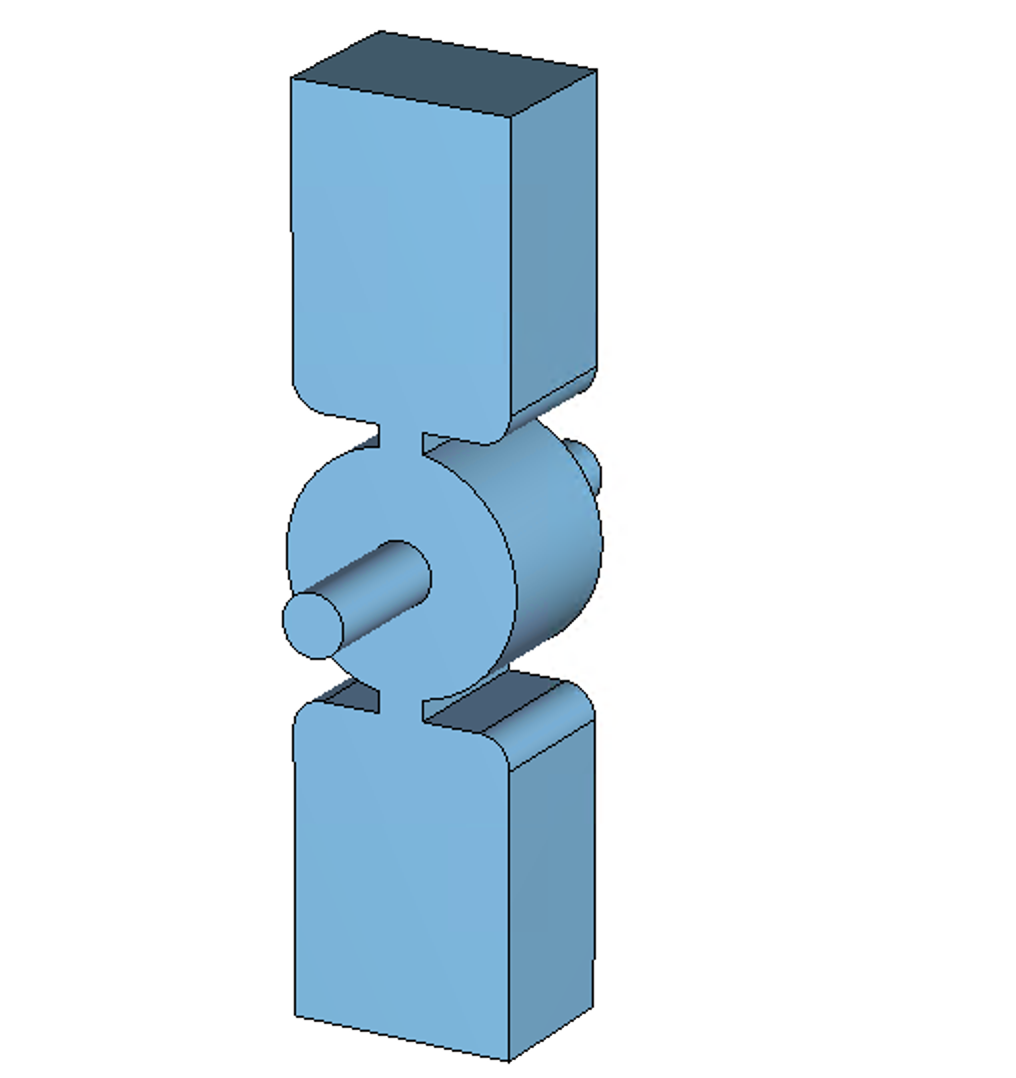}}
\subfloat[][4-port coupler.\label{subfig:4_Port}]{\includegraphics[width=0.16\textwidth]{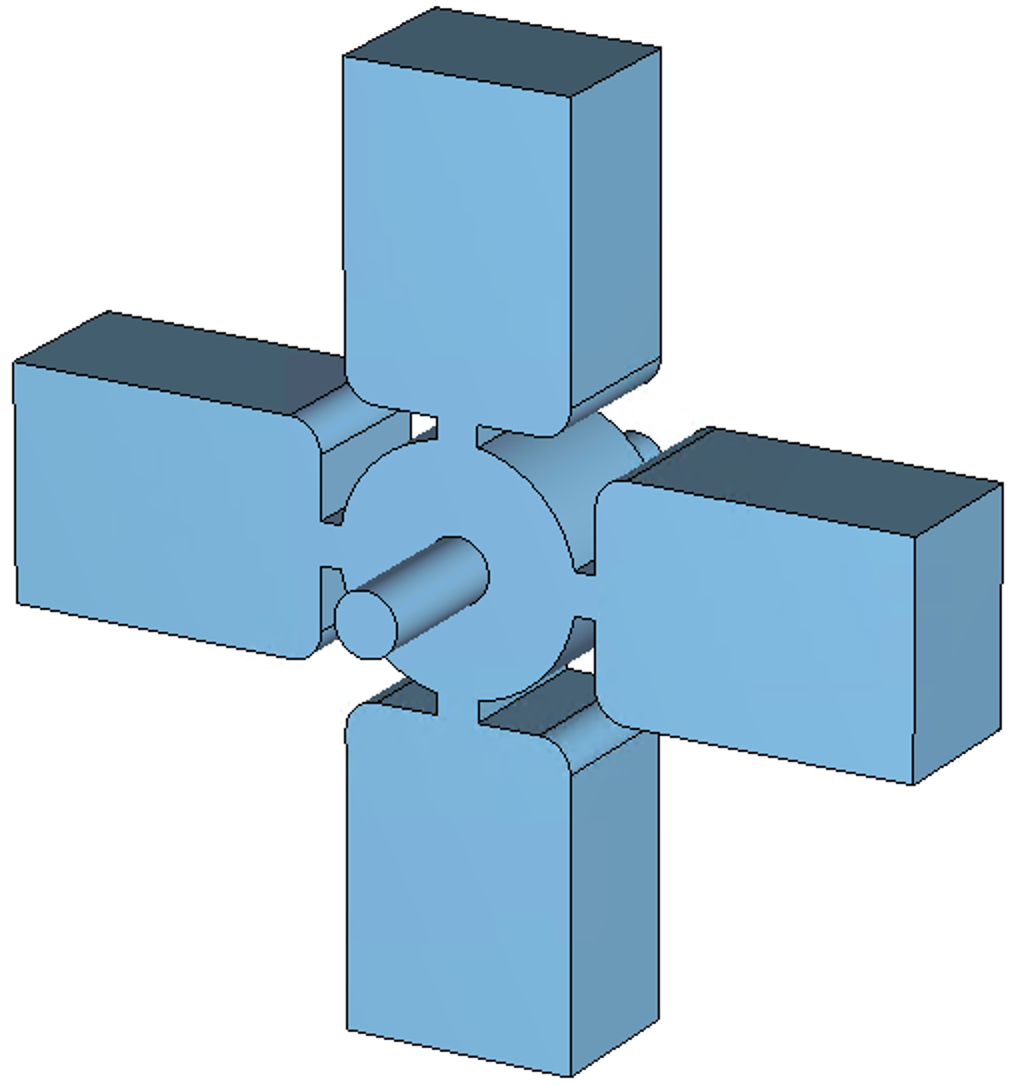}}
\caption{Pillbox cavities incorporating beam pipes and different $N$-port couplers.}\label{fig:N_Port}
\end{figure}

Here we analyse the \qty{3}{GHz} TM$_{010}$  modes supported by rf cavities with a cell length of $L_c=c/2f_l\simeq$~\qty{50}{mm} and beam pipes of radius $a~=~$\qty{10}{mm} and length $L_p~=~$\qty{50}{mm}. Figure~\ref{fig:N_Port} shows the design of such pillbox cavities with 1-, 2-, and 4-port couplers incorporated. These structures were constructed in \textsc{CST} and simulated with simultaneous excitations of all rf ports (with equal amplitude and phase). To critically couple the cavities, slot-widths of \qty{15.7}{mm}, \qty{14.6}{mm}, and \qty{13.2}{mm} were required for the 1-, 2-, and 4-port coupler designs respectively.
 
For each design, the TM$_{010}$-like mode was solved and the transverse multipolar components of the mode measured by integrating the electric field as per Eq.~\ref{eq:Pz_Change} and subsequently decomposing the longitudinal momentum into the form given by
\begin{equation} \label{eqn:DeltaPz_Free}
    \Delta p_{z}(r,\theta) = \sum_{m=0}^\infty\tilde{\gamma}_mr^m\cos{(m\theta+\phi_m}).
\end{equation}
This expression follows from Eq.~\ref{eq:Pz_Azi_Final2}, where, for simplicity, all constants have been absorbed into the coefficient $\gamma_m$. Table~\ref{tab:PortMultipoles} shows the results of this decomposition for the different cavity designs with $<~$\qty{300000}{} mesh cells. We see that the dominant transverse multipole for the $N$-port coupler design is the $\tilde{\gamma}_N$ multipole.

\begin{table}[h!]
\caption{\label{tab:PortMultipoles}
Momentum multipolar components in the \qty{3}{GHz} TM$_{010}$-like mode supported by a pillbox with beam pipe and $N$-slot couplers.}
\begin{ruledtabular}
\begin{tabular}{c|cccc}
& $\tilde{\gamma}_1/\tilde{\gamma}_0$ & $\tilde{\gamma}_2/\tilde{\gamma}_0$ & $\tilde{\gamma}_3/\tilde{\gamma}_0$ & $\tilde{\gamma}_4/\tilde{\gamma}_0$ \\
\hline
1-port  & $0.0119$ & $0.0035$ & $0.0012$ & $0.0004$  \\ 
2-port  & $<0.0001$ & $0.0069$ & $<0.0001$ & $0.0009$ \\
4-port  & $<0.0001$ & $0.0001$ & $<0.0001$ & $0.0019$ \\
\end{tabular}
\end{ruledtabular}
\end{table}

\subsection{Multipole-free design} \label{sec:MultipoleFreeMethod}
 
Adding ports to negate the transverse multipoles is not only cumbersome but also impractical for removing the octupole component due to the limited space, which typically allows no more than a 4-port configuration. An alternative is to retain a 1-port coupler and use the AMM to minimise all transverse multipoles, thus creating a \textit{multipole-free} accelerating structure. The following iterative process is used to achieve this:
\begin{itemize}
    \item The cavity at iteration step $i$ (with $i=0$ as the circular pillbox) is constructed in \textsc{CST}.
    \item The slot-width of the 1-port coupler is optimised for critical coupling (S$_{11}~<~$\qty{-40}{dB}).
    \item The cavity is scaled such that the TM$_{010}$ mode resonates at \qty{3}{GHz}.
    \item The TM$_{010}$ mode is solved, and the longitudinal electric field is exported along a cylinder of radius $R$ and length $2(L_c/2+L_p)$.
    \item The voltage $V_z(R,\theta)=\int^{L_c/2+L_p}_{-(L_c/2+L_p)} E_z(R,\theta,z)dz$ is calculated and decomposed into Bessel functions as $V_z=\tilde{\nu}_mJ_m(k_lR)\cos{(m\theta+\phi_m)}$.
    \item The correcting multipoles for this iteration are calculated using $\tilde{g}_m^{(i)} = \tilde{\nu}_m/J_0(k_lR)$.
    \item A new cavity shape is calculated by solving Eq.~\ref{eq:AMMBC} with $\tilde{g}_m=\sum_i-\tilde{g}_m^{(i)}$ (where the sum is over the number of iterations $i$) and the associated $\phi_m$.
\end{itemize}
It should be noted that the non-trivial geometry of the cavity and coupler can be represented by Fourier decomposing the boundary into a sufficiently large number of components. Consequently, accurate reconstruction requires decomposing the on-axis voltage into Bessel-based multipolar components of sufficiently high order to capture the highest azimuthal variations. In practice, however, numerical noise in the field data and diminishing contributions of higher-order multipoles make it computationally expensive to resolve beyond a certain order, typically the dodecupole. In the example presented here, the boundary condition is solved using components up to the octupole, which, as presented below, is sufficient to yield a multipole-free accelerating structure.

\begin{figure*}[tbh!]
\includegraphics[width=0.75\textwidth]{Figures/LegendField.png}\\
\subfloat[][1-port coupler.\label{subfig:SingleField}]{\includegraphics[width=0.35\textwidth]{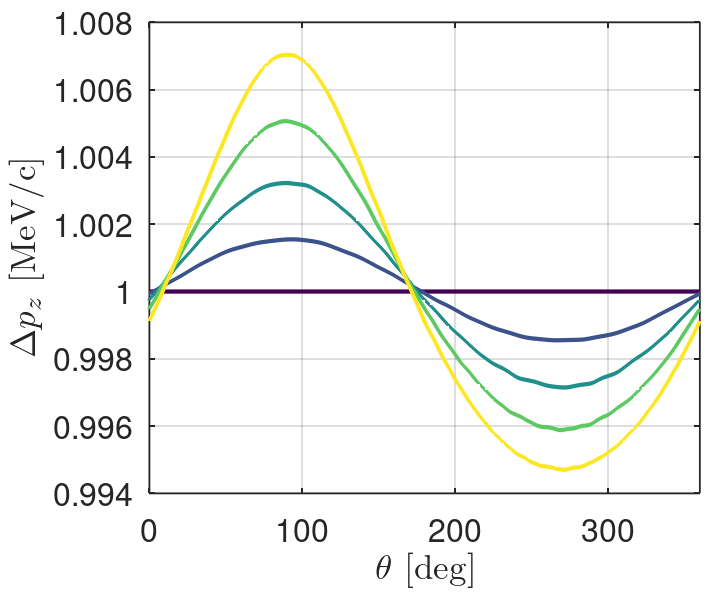}}
\subfloat[][2-port coupler.\label{subfig:DualField}]{\includegraphics[width=0.35\textwidth]{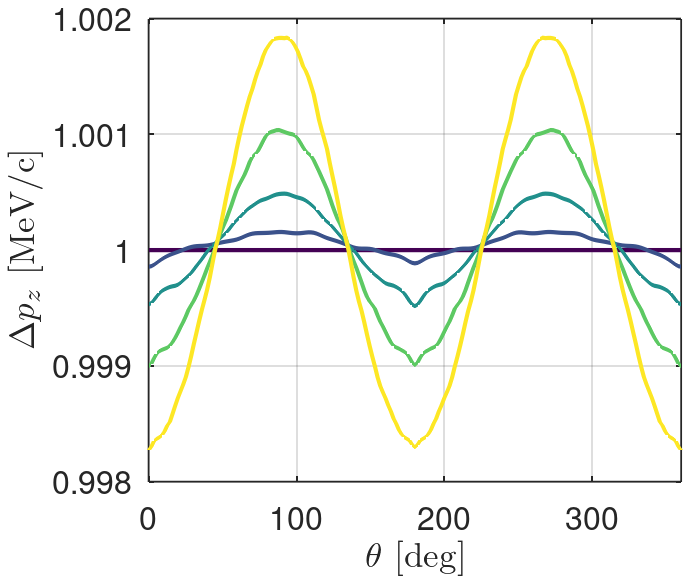}}
\\
\subfloat[][4-port coupler.\label{subfig:QuadField}]{\includegraphics[width=0.36\textwidth]{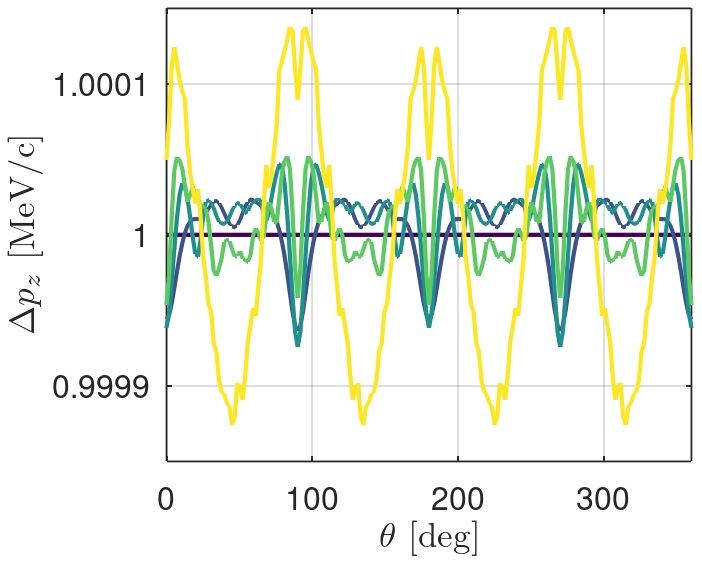}}
\subfloat[][Multipole-free.\label{subfig:AMMField}]{\includegraphics[width=0.36\textwidth]{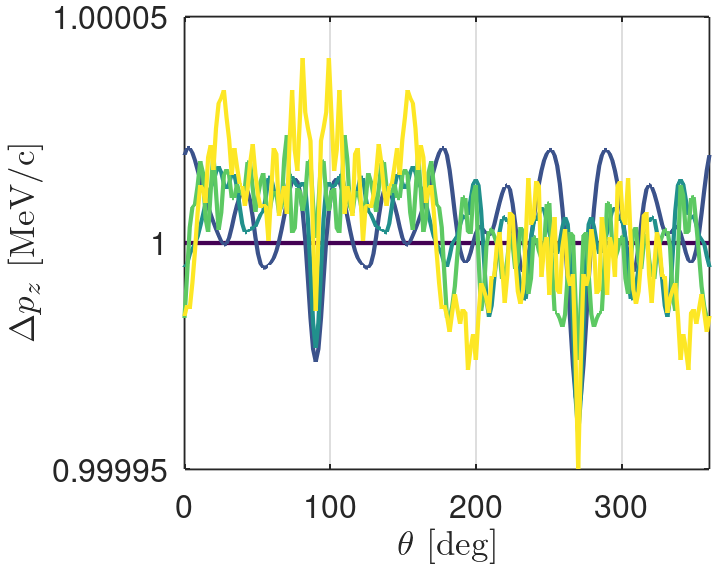}}
\caption{Change in longitudinal momentum of a rigid, ultra-relativistic, parallel particle traversing \qty{3}{GHz} TM$_{010}$-like modes in different cavity configurations: (a) circular pillbox with 1-port coupler, (b) with 2-port coupler, (c) with 4-port coupler and (d) multipole-free accelerating structure with 1-port coupler. The electromagnetic field is normalised such that the on-axis change in momentum is \qty{1}{MeV/c}.}\label{fig:FieldComparison}
\end{figure*}

After five iterations of applying the above method to the single-slot \qty{3}{GHz} TM$_{010}$ pillbox cavity, a multipole-free accelerating structure (with a slot-width of \qty{16.7}{mm}) was achieved, where all transverse multipoles were measured as $\tilde{\gamma}_m/\tilde{\gamma}_0<0.0001$. Figure~\ref{fig:FieldComparison} compares the multipole-free design to the $N$-port pillbox designs by plotting $\Delta p_z$, calculated by integrating the longitudinal electric field per Eq.\ref{eq:Pz_Change}, as a function of $\theta$ at different radii when $V_{z,0}$~=~\qty{1}{MV} (corresponding to an average gradient of \qty{20}{MV/m}). The finite mesh resolution and the field map sampling with a step-size of \qty{0.25}{mm} introduce some noise and error. Nevertheless, the dominant dipole, quadrupole, and octupole components are clearly visible in the 1-port (Fig.~\ref{subfig:SingleField}), 2-port (Fig.~\ref{subfig:DualField}), and 4-port (Fig.~\ref{subfig:QuadField}) structures respectively. Additionally, the decreasing influence of the transverse multipoles on $\Delta p_z$ as $N$ increases is also visible. For instance, the dipole component in the 1-port coupler causes a maximum perturbation of \qty{0.7}{\%} at $r~=~$\qty{8}{mm}, the quadrupole component in the 2-port causes \qty{0.2}{\%}, and the octupole component in the 4-port causes \qty{0.01}{\%}.  

Figure~\ref{subfig:AMMField} shows $\Delta p_z$ for the 1-port, multipole-free cavity design. The transverse multipole components are reduced to the order of the noise floor ($<$~\qty{0.005}{\%}) corresponding to the dipole, quadrupole, sextupole, and octupole components being smaller than $\gamma_m/\gamma_0~<~$\qty{0.00005}{}. Increasing the mesh sampling size and reducing the field-map step-size could further reduce these components, at the expense of increased computation time.

\begin{figure}[h!]
\subfloat[][Pillbox.\label{subfig:Coupler_Azi}]{\includegraphics[width=0.16\textwidth]{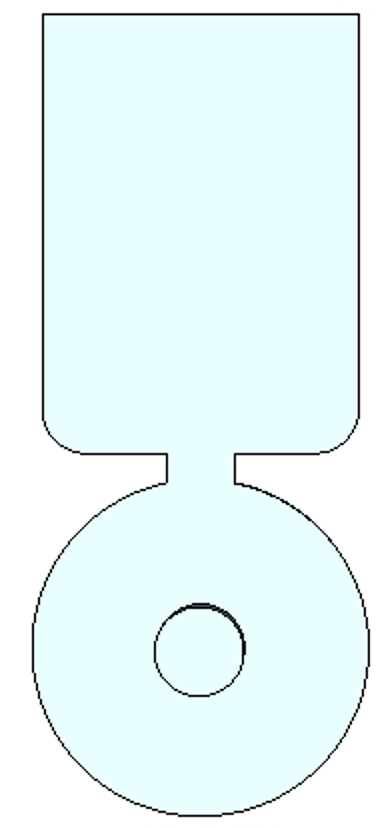}}
\subfloat[][Multipole-free.\label{subfig:Coupler_Azi2}]{\includegraphics[width=0.1525\textwidth]{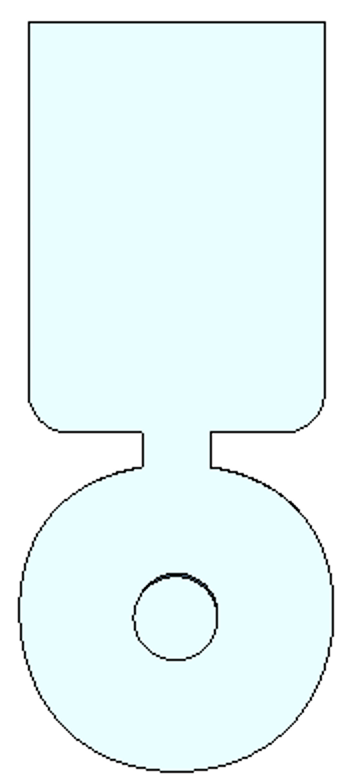}}
\\
\subfloat[][Cross-section comparison.\label{subfig:ShapeCompare}]{\includegraphics[width=0.28\textwidth]{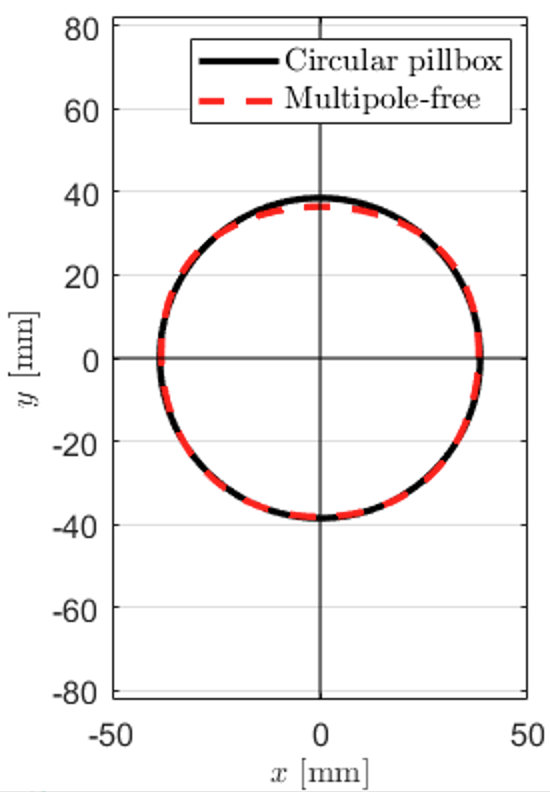}}
\subfloat[][Cross-section difference.\label{subfig:ShapeDiff}]{\includegraphics[width=0.21\textwidth]{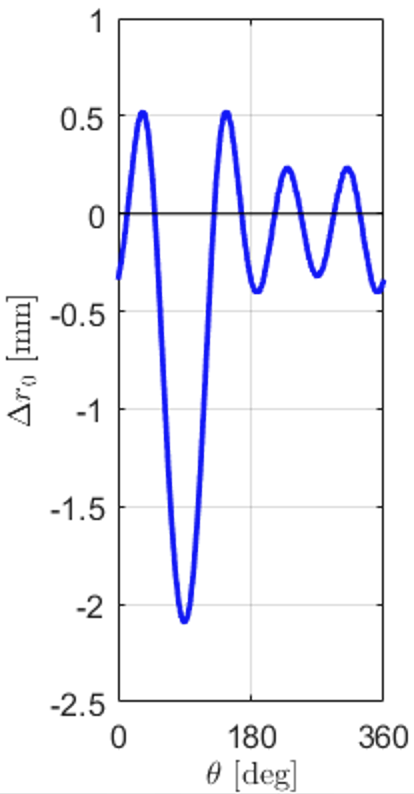}}
\caption{Circular pillbox with 1-slot coupler (a) and multipole-free cavity with 1-slot coupler (b). Cross-section comparisons in Cartesian coordinates are shown in (c), and radial differences as a function of angle in (d).}\label{fig:ShapeComparison}
\end{figure}

Figure~\ref{fig:ShapeComparison} compares the azimuthal cross-section of the multipole-free design with that of the circular pillbox. We see that the multipole-free design is compressed relative to the pillbox at $\theta~=$~\ang{90}, where the 1-port coupler connects. The difference in the cross-sections varies by up to \qty{1.3}{mm} (\qty{3.4}{\%} of the pillbox cavity radius), which is well-within manufacturing tolerances.

The minimisation of the transverse multipoles in the multipole-free design reduces the change in transverse momentum, as described by Eq.~\ref{eq:Pperp_Change}. This results in two main benefits of designing multipole-free accelerating structures: better minimisation of the transverse multipoles and the elimination of the need for additional ports, which would otherwise add complexity to the design process, increase manufacturing time, and require extra space and material.

\section{Uniform Beam} \label{sec:UniformBeam}

In addition to removing unwanted multipoles, the AMM can also be used to introduce desired multipolar components. In this section, we present an example application of using the AMM to design an rf cavity that transforms the transverse spatial distribution of a beam from Gaussian to uniform.

\subsection{Uniformisation with multipole magnets}

Reference~\cite{UniformBeam} presents a method for uniformising the transverse beam profile by utilising the nonlinear-focusing forces of a multipole magnet. The approach requires a multipole magnet composed of an octupole with integrated field strength
\begin{equation} \label{eqn:IdealOct}
    K_4 = \frac{1}{2\varepsilon\beta_0}\frac{1}{\beta_0\tan\Phi},
\end{equation}
and higher even-order $m$-poles defined by the recursive relationship
\begin{equation} \label{eq:IdealHigher}
    K_{m+2} = -\frac{m-1}{\varepsilon\beta_0}K_{m}, \;\; (m~=~4,~6,~8,~\cdots)\\,
\end{equation}
where $\varepsilon$ is the geometric rms emittance of the beam, $\beta_0$ is the beta function at the magnet position $0$, and $\Phi$ is the betatron phase advance from the magnet to the target position $t$ where the beam becomes uniform. The magnet must be placed at a point where the beam size in one direction is significantly larger than the other --- for analysis here, let us assume this element is located at a point where the beam has a finite $x$ dimension and negligible $y$ dimension. The multipole magnet uniformises the beam to a horizontal distribution with a transverse diameter given by
\begin{equation}\label{eqn:Target_r}
    2r_t = \sqrt{2\pi}\sqrt{\varepsilon\beta_t}|\cos{\Phi}|,
\end{equation}
where $\beta_t$ is the beta function at the target position.

To uniformise the beam in both horizontal and vertical directions,  a second multipole magnet is required. This magnet must be positioned where the vertical envelope of the beam is significantly larger than the horizontal one. 

\subsection{Theory of uniformisation with rf cavities}
The multipole magnet transforms the Gaussian beam into a uniform distribution by non-linearly changing the transverse momentum of the particles across different horizontal positions, perfectly folding the tails of the Gaussian distribution onto each other at position $z_t$. The multipolar expansion of the required magnetic field described in Eqns.~\ref{eqn:IdealOct}~-~ \ref{eq:IdealHigher} is
\begin{equation} \label{eq:IdealMultiB}
    \vec{B}(r,\theta) = \frac{P_\textrm{ref}}{q}\sum^\infty_{m=4}\frac{K_{m}}{L_c}\frac{r^{m-1}}{(m-1)!}\begin{pmatrix}\sin{m\theta} \\ \cos{m\theta}\end{pmatrix}_{(r,~\theta)},
\end{equation}
where $P_\textrm{ref}$ is the momentum of the reference particle and $L_c$ is the length of the magnet. The resulting change in transverse momentum is calculated by integrating Eq.~\ref{eq:IdealMultiB} using the Lorentz force, giving
\begin{equation} \label{eq:Uniform_P_Sum}
    \Delta\vec{p}_\perp(r,\theta) = P_\textrm{ref}\sum^\infty_{m=4}K_{m}\frac{r^{m-1}}{(m-1)!}\begin{pmatrix}-\cos{m\theta} \\ \sin{m\theta}\end{pmatrix}_{(r,~\theta)}.
\end{equation}

The multipolar composition needed to uniformise a beam using an rf cavity can be calculated by equating the multipolar expansion in  Eq.~\ref{eq:Uniform_P_Sum} to Eq.~\ref{eq:Pperp_Azi_Final} which gives the change in transverse momentum of a particle traversing a TM$_{\{m\}10}$ mode. The multipolar bore fields of the needed mode relate to the integrated multipole magnet strengths as
\begin{equation}
   \tilde{G}_m = \frac{P_\textrm{ref}}{q} \frac{\omega_l}{L_c} \frac{k_lL_c/2}{\sin{\left(k_lL_c/2\right)}}\frac{1}{\sin\psi} \frac{a^m}{m!}K_m.
\end{equation}
By re-expressing this using Eq.~\ref{eq:BoreRel}, we obtain the following relations for the rf cavity multipolar strengths required to perfectly uniformise a beam 
\begin{equation} \label{eqn:Idealg_m}
   \tilde{g}_m = \frac{P_\textrm{ref}}{q} \frac{\omega_l}{L_c} \frac{k_lL_c/2}{\sin{\left(k_lL_c/2\right)}}\frac{1}{\sin\psi} \frac{a^m}{J_m(k_la)m!}K_m.
\end{equation}

\subsection{Beam dynamics simulations with rf cavities}

\begin{figure}[b!]
    \centering
    \includegraphics[width=\linewidth]{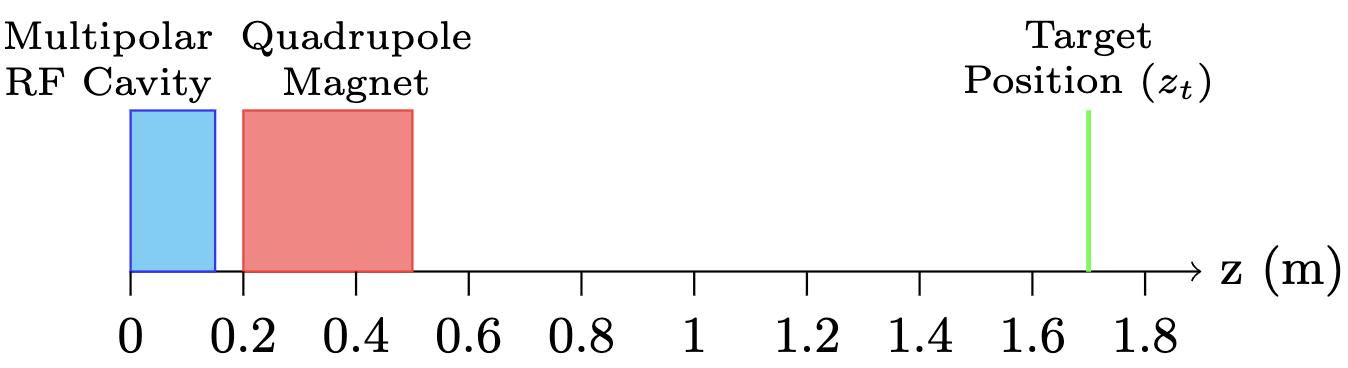}
    \caption{Illustration of the beam testline used in tracking simulations.}
\label{fig:Testline}
\end{figure}

To investigate this application of uniformising beams with rf cavities, we undertook beam dynamics simulations based on a testline similar to the one analysed in Ref.~\cite{UniformBeam}. The schematic layout of this testline is shown in Fig.~\ref{fig:Testline}. An rf cavity of length \qty{0.15}{m} is placed at $z~=~0$, followed by a quadrupole magnet of length \qty{30}{cm} and integrated strength $K_2~=~$\qty{4.25}{m^{-1}} placed at $z~=~$\qty{20}{cm}.  The rf cavity is designed for \qty{3}{GHz} operation, with a cell length $L_c=$~\qty{5}{cm} and beam pipes of radius $a~=~$\qty{50}{mm} and length $L_p~$=~\qty{5}{cm}.

We tracked \qty{200000}{} particles through the testline with an initial momentum $P_0$~=~\qty{10}{MeV} and using the same initial beam conditions used in Ref.~\cite{UniformBeam}: $\beta_0$~=~\qty{15}{m}, $\alpha_0$~=~\qty{15}{}, and $\varepsilon$~=~\qty{21}{mm\;mrad}. The phase advance to the target position at $z_t~=~$\qty{1.7}{m} is $\Phi~=~$\qty{0.0678}{} and, the $\beta$-function at the target position is $\beta_t~=~$\qty{223}{m}. As a result, we expect the ideal multipolar field distribution to generate a uniform beam of radius $r_t~=~$\qty{85.6}{mm} by Eq.~\ref{eqn:Target_r}.


\subsubsection{Pillbox cavities}

\begin{figure}[b]
\subfloat[][Cavity.\label{subfig:TM_410_Cav}]{\includegraphics[width=0.21\textwidth]{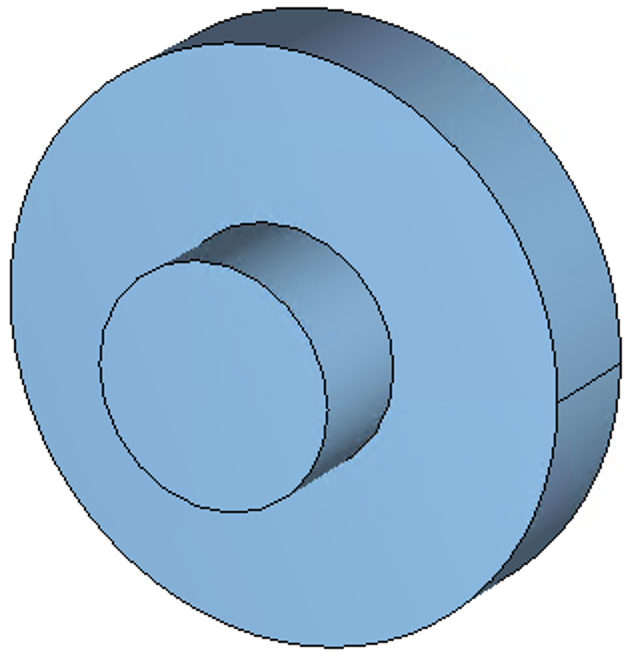}}
\subfloat[][Mode.\label{subfig:TM_410}]
{\includegraphics[width=0.28\textwidth]{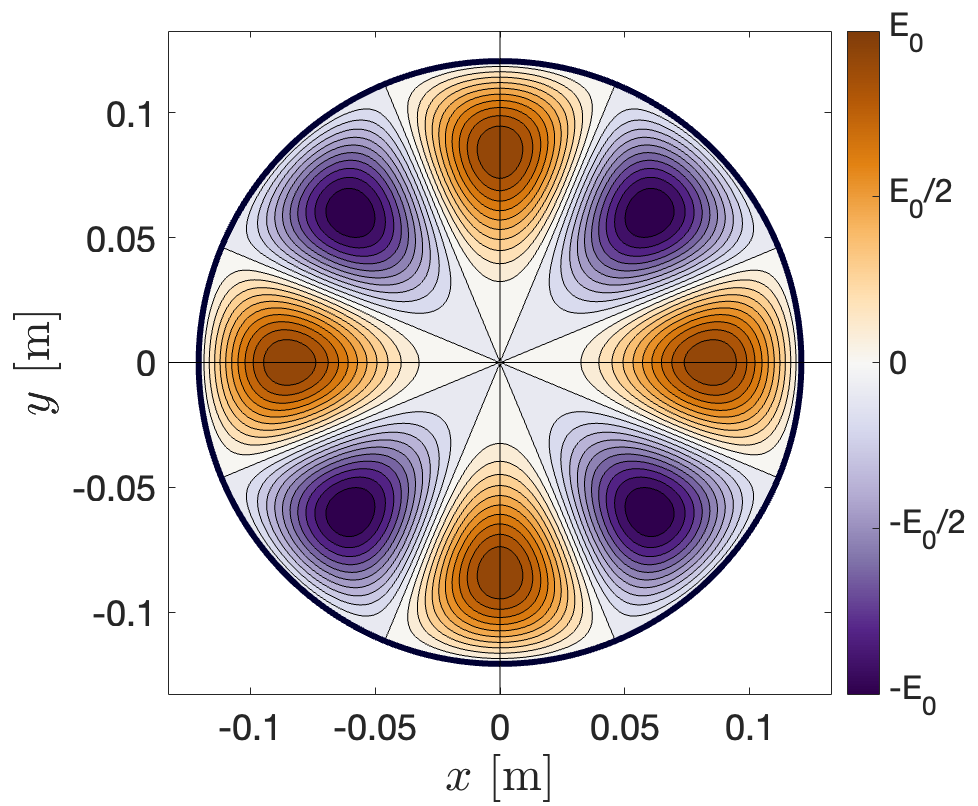}}
\caption{Pillbox cavity designed to support a \qty{3}{GHz} TM$_{410}$ mode (a) and the corresponding contour plot of $E_z$ at the centre of the cavity (b).}\label{fig:UniTM410}
\end{figure}

As a first test, we placed a pillbox cavity operating in a TM$_{410}$ mode into the testline. The pillbox and its mode are shown in Figure~\ref{fig:UniTM410}, with the electromagnetic field solved in \textsc{CST}. The field map within the cylinder of radius $r~=~a$ was imported into \textsc{RF-Track} with a step-size of \qty{0.3}{mm}.  

\begin{figure}[t]
    \centering
    \includegraphics[width=\linewidth]{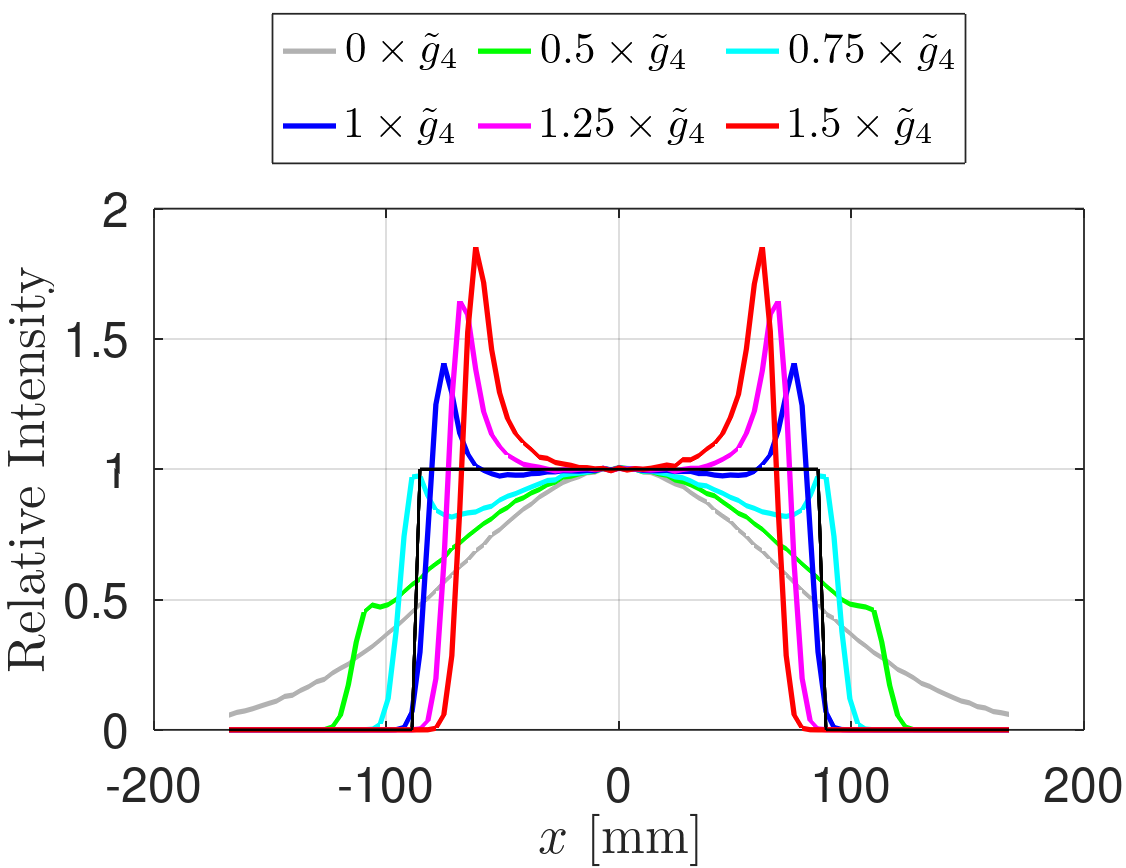}
    \caption{Horizontal beam distribution at the target position $z_t$ in the testline with a pillbox cavity operating in a TM$_{410}$ mode, where the different octupolar strengths are indicated in the legend with $\tilde{g}_4$~=~\qty{108}{MV/m}. The black line represents the ideal uniform distribution.}
    \label{fig:UniformOctupole}
\end{figure}

The magnitude of the ideal octupole component for beam uniformisation is calculated using Eq.~\ref{eqn:Idealg_m} as $\tilde{g}_4$~=~\qty{108}{MV/m}. Figure~\ref{fig:UniformOctupole} compares the transverse intensity distribution at the target location for different octupole field strengths of: $0\times\tilde{g}_4$, $0.75\times\tilde{g}_4$, $1\times\tilde{g}_4$, $1.25\times\tilde{g}_4$, and $1.5\times\tilde{g}_4$. We observe that a lower octupole magnitude is insufficient to fold the tails of the Gaussian distribution, resulting in no uniform region. On the other hand, a larger magnitude introduces a uniform region at the centre of the distribution, but it is too strong and overfolds the tails to create horns at the edges of the distribution.

Despite the horns, the TM$_{410}$ mode can uniformise the beam effectively. For the $1\times\tilde{g}_4$ case, $|x|~\lesssim~$\qty{60}{mm} contains \qty{71}{\%} of the total particles and is uniform to within $<~$\qty{4}{\%}. 

In the pillbox cavity supporting this mode, the maximum longitudinal electric field at the bore gap is $\qty{16.4}{MV/m}$, while within the bulk of the structure it reaches $\qty{43.0}{MV/m}$. This peak field remains within practical breakdown limits and, as per Eq.~\ref{eqn:Idealg_m}, this peak field could be reduced by effectively increasing the cavity length by, for example, constructing it out of a series of coupled cells.

\begin{figure}[b]
    \centering
    \includegraphics[width=0.9\linewidth]{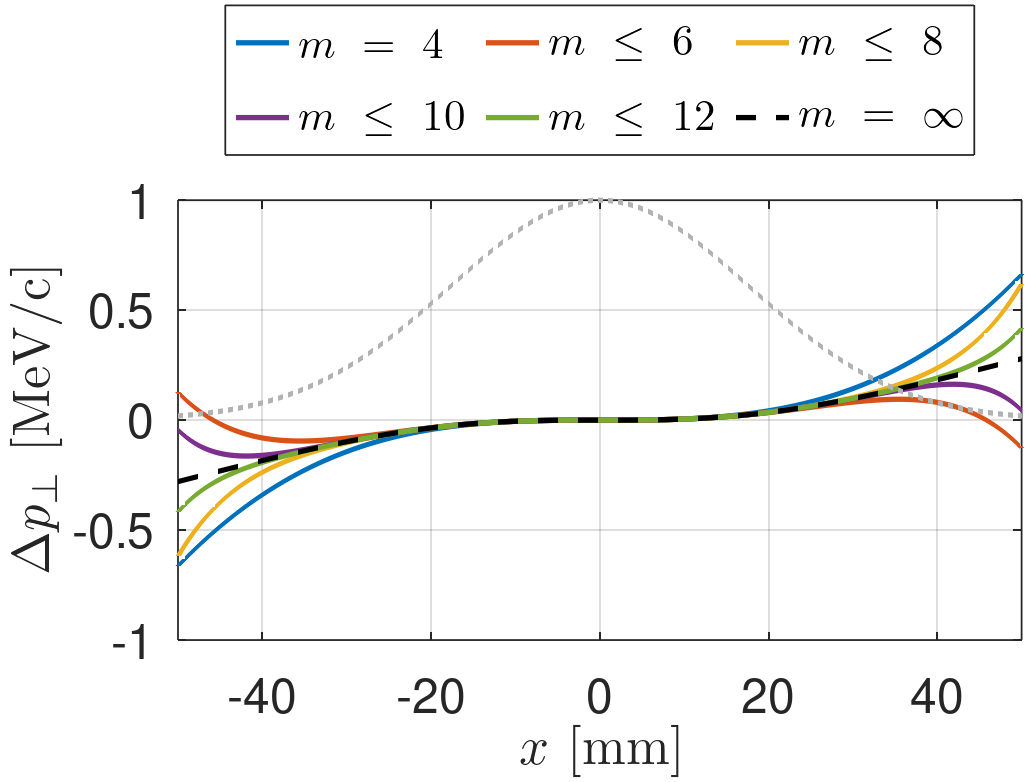}
    \caption{Transverse momentum change imparted by an ideal, thin uniformising element with multipoles up order $m$. The dashed grey line represents the Gaussian distribution of the initial beam.}
    \label{fig:UniformIdeal}
\end{figure}

To compensate for the over-folding of the beam tails, higher order multipoles must be included. To demonstrate this, Fig.~\ref{fig:UniformIdeal} shows the change in momentum imparted by the ideal multipolar element calculated using Eq.~\ref{eq:Uniform_P_Sum}, as successively higher orders of $m$ are included. We observe that including additional multipole components improves convergence towards the ideal transverse momentum profile, especially in the tails of the Gaussian distribution. 

It is important to note, however, that the ideal multipolar values calculated from Eqns.~\ref{eqn:IdealOct}~-~\ref{eq:IdealHigher} are based on the assumption of a thin multipolar element. In contrast, the rf cavities considered here are thick structures traversed by a converging beam ($\alpha_0$~=~15). As such, the theoretically predicted values should be treated as approximate rather than exact, estimates for the optimal multipolar strengths required to uniformise the beam. 



\begin{figure}[t!]
\subfloat[][Cavity.\label{subfig:TM_610_Cav}]{\includegraphics[width=0.21\textwidth]{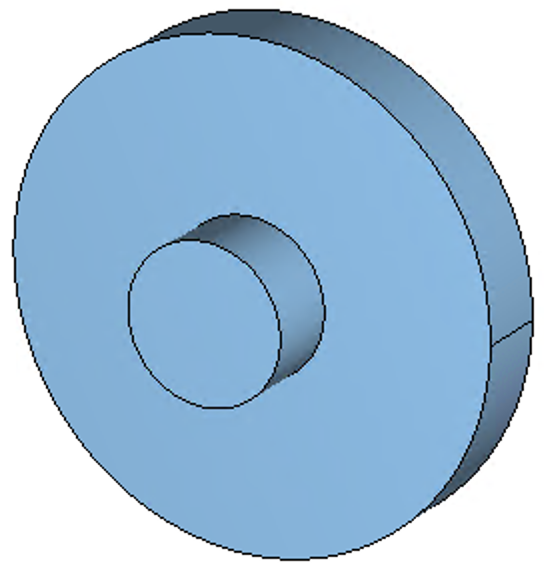}}
\subfloat[][Mode.\label{subfig:TM_610}]
{\includegraphics[width=0.28\textwidth]{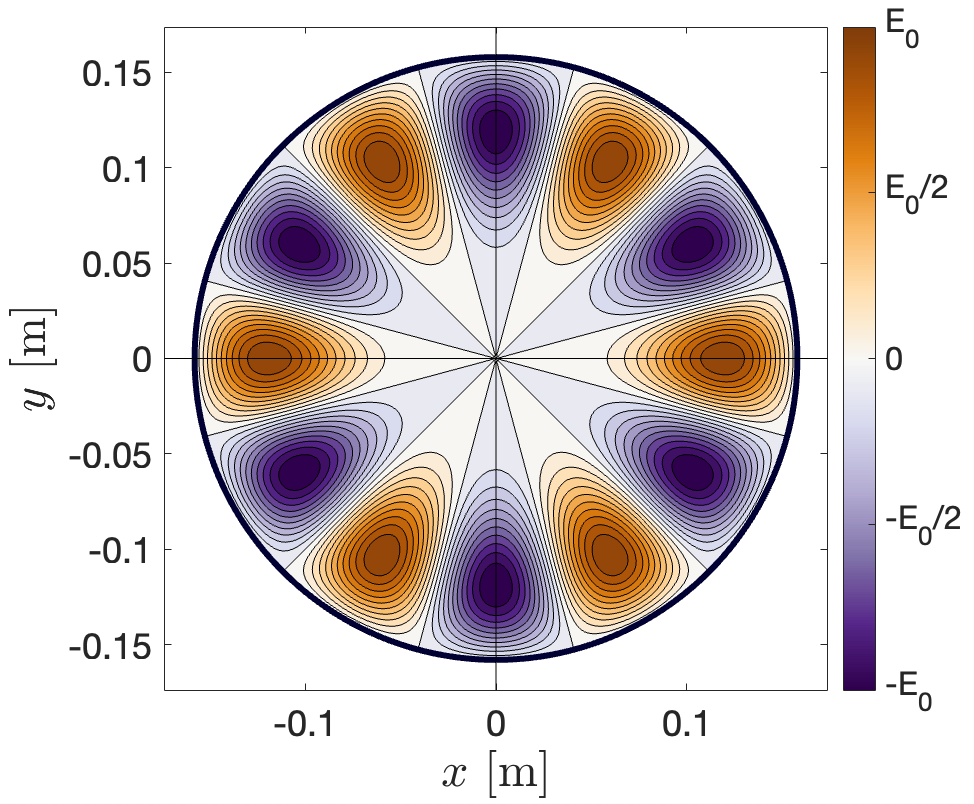}}
\caption{Pillbox cavity designed to support a \qty{3}{GHz} TM$_{610}$ mode (a) and the corresponding contour plot of $E_z$ at the centre of the cavity (b).}\label{fig:UniTM610}
\end{figure}

%
Bearing this in mind, the next multipole to try is a dodecapole which, from Eq.~\ref{eqn:Idealg_m}, we estimate the required strength as $\tilde{g}_6$~=~\qty{-891}{MV/m}. To realise this component, we designed a pillbox cavity that operates in a $\qty{3}{GHz}$ TM$_{610}$ mode, as shown in Fig.~\ref{fig:UniTM610}. 

We superimposed this field with the TM$_{410}$  field at the same position in the testline. (This is unphysical because the two pillbox cavities have different radii and cannot physically occupy the same space. This anticipates the use of the AMM to resolve this limitation). This setup provides two tunable parameters, $\tilde{g}_4$ and $\tilde{g}_6$, which we vary to generate the most uniform beam possible. 

\begin{figure}[b]
    \centering
    \includegraphics[width=0.8\linewidth]{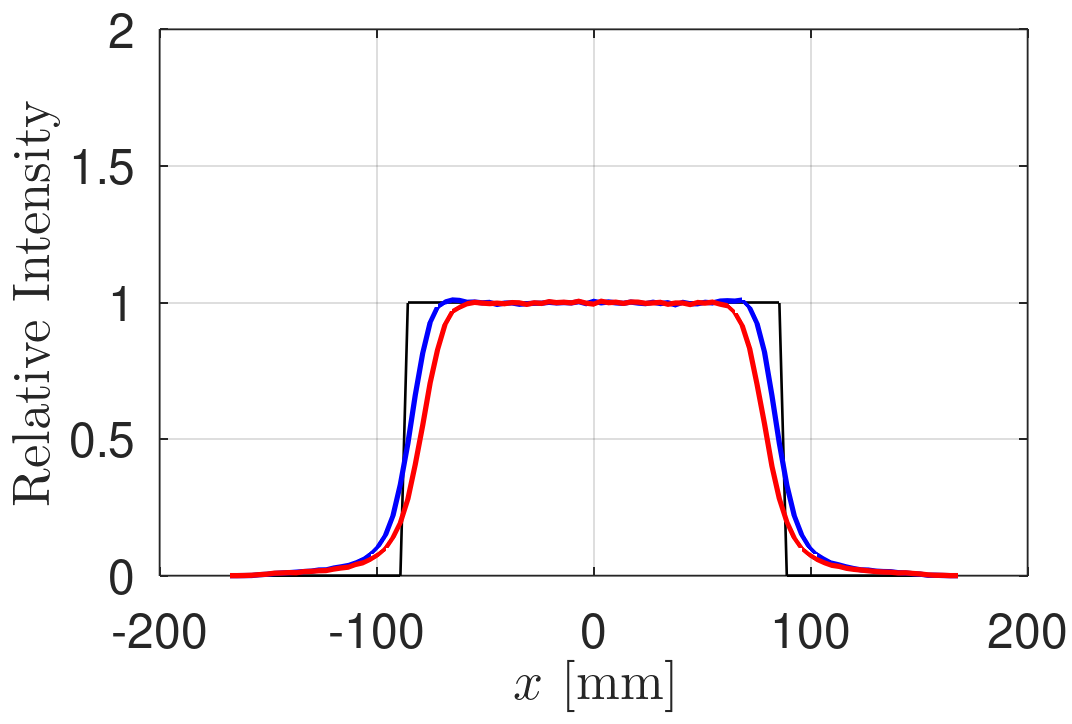}
    \caption{Horizontal beam distribution at the target position $z_t$ in the testline with a multipolar field consisting of a TM$_{410}$ superimposed with a TM$_{610}$ mode (blue), and for an azimuthally modulated cavity supporting a TM$_{\{4,6\}10}$ (red). The black line represents the ideal uniform distribution.}
    \label{fig:UniTM4610}
\end{figure}

Using an optimisation process that minimises the residual sum of squares between the tracked distribution and the ideal uniform distribution of radius $r_t$, we found the optimal multipolar values to be $\tilde{g}_4$~=~\qty{151}{MV/m} and $\tilde{g}_6$~=~\qty{-1006}{MV/m} (corresponding to a ratio of $\tilde{g}_6/\tilde{g}_4$=\qty{-6.68}{}). The resulting distribution at the target position is shown in Fig.~\ref{fig:UniTM4610} (blue line). We see that addition of the dodecapole field effectively smooths out the horns present in the TM$_{410}$-only case, producing a distribution that contains \qty{82}{\%} of the particles and is uniform to within \qty{1}{\%} in the region $~|x|~\lesssim~$\qty{70}{mm}.

\subsubsection{Azimuthally modulated cavities}

\begin{figure}[b]
\subfloat[][Cavity.\label{subfig:UniformCav1}]{\includegraphics[width=0.17\textwidth]{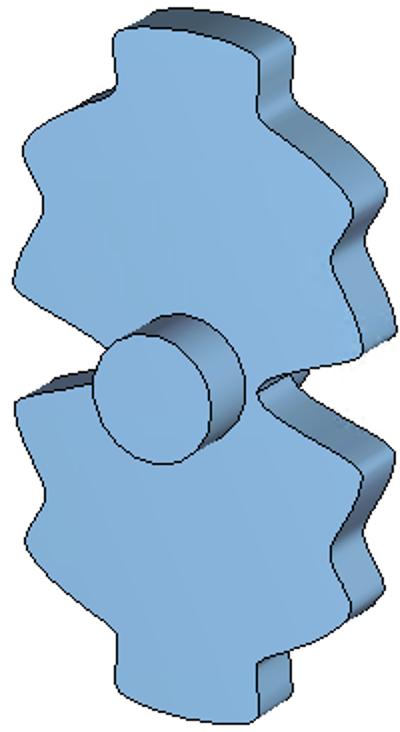}}
\subfloat[][Mode.\label{subfig:UniformCav2}]{\includegraphics[width=0.32\textwidth]{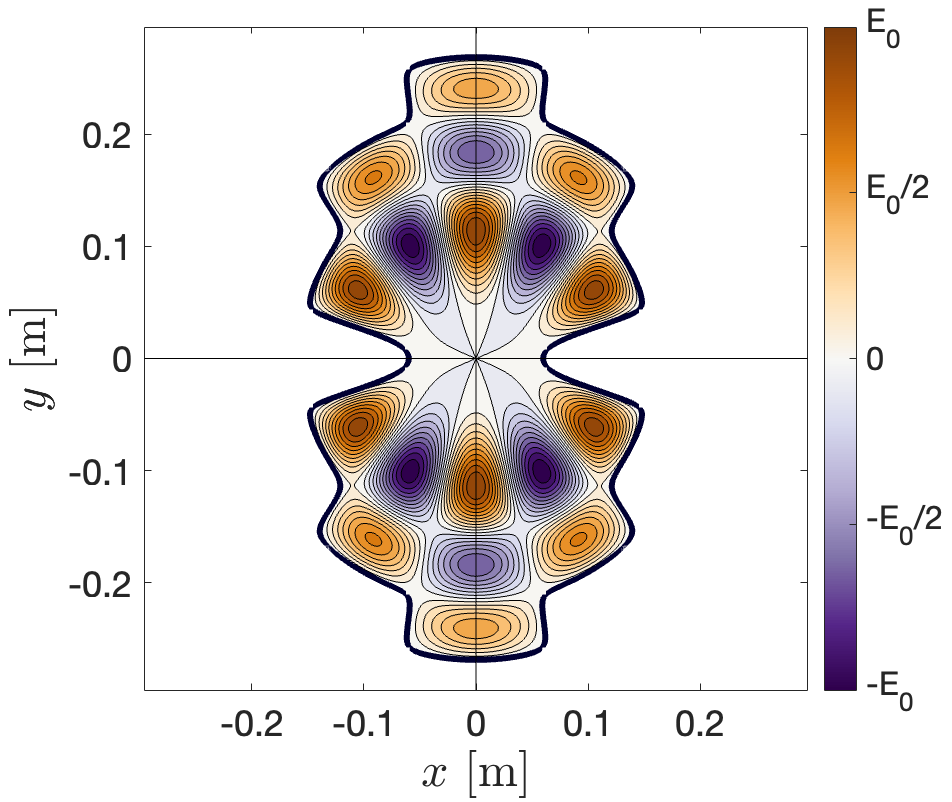}}
\caption{Azimuthally modulated cavity designed to support a \qty{3}{GHz} TM$_{\{4,6\}10}$ mode with $\tilde{g}_6/\tilde{g}_4$=\qty{-7.10}{} (a) and the corresponding contour plot of $E_z$ at the centre of the cavity (b).}\label{fig:Uniform_RF}
\end{figure}

Taking the previous result, we solve Eq.~\ref{eq:AMMBC} with the optimised ratio $\tilde{g}_6/\tilde{g}_4$=\qty{-6.68}{} to design an azimuthally modulated cavity that supports a \qty{3}{GHz} TM$_{\{4,6\}10}$ mode. The resultant structure and solved for mode are shown in Fig.~\ref{fig:Uniform_RF}. 

In the solved mode, we measure  $\tilde{g}_6/\tilde{g}_4$~=~\qty{-7.10}{} which is \qty{5}{\%} larger than the design value. By varying the field strength of the mode (with only one free parameter, since $\tilde{g}_6/\tilde{g}_4$ is fixed), we find that the optimal values for achieving a uniform beam are $\tilde{g}_4$~=~\qty{135}{MV/m} and $\tilde{g}_6$~=~\qty{-958}{MV/m}. In this TM$_{\{4,6\}10}$ mode, the maximum longitudinal electric field at the bore gap is $\qty{14.8}{MV/m}$, while within the bulk of the structure it reaches $\qty{159}{MV/m}$.

The resulting uniform distribution at the target location is shown in Fig.~\ref{fig:UniTM4610} (red line), where approximately \qty{74}{\%} of the particles are contained within the region $~|x|~\lesssim~$\qty{60}{mm}, and the distribution is uniform to within $\qty{0.3}{\%}$. The TM$_{\{4,6\}10}$ mode provides comparable performance to the superposition of the TM$_{410}$ and TM$_{610}$ modes, smoothing out the horns in the distribution and even producing a more uniform profile, although over a slightly smaller region. These differences arise due to small discrepancies between the desired and achieved field. Specifically, there is a $<6\%$ difference between the targeted and measured multipolar ratio ($\tilde{g}_8/\tilde{g}_6$~=~-6.68 vs -7.10), along with minor residual monopole and quadrupole components. 

An iterative design process, similar to that described in Sec.~\ref{sec:MultipoleFreeMethod}, could be used tune the multipolar components and compensate for this small discrepancy. Then, to further uniformise the beam, additional higher-order multipolar components need to be added. As guided by Eq.~\ref{eqn:Idealg_m}, these components are expected to be on the order of $\tilde{g}_8/\tilde{g}_4~=~$\qty{122}{}, $\tilde{g}_{10}/\tilde{g}_4~=~$\qty{-2.60e3}{}, and $\tilde{g}_{12}/\tilde{g}_4~=~$\qty{7.27e4}{}. Further investigation, and the design of such a structure for implementation in an actual beamline, is beyond the scope of this paper. Nevertheless, the example highlighted in this section of uniformising a beam with an rf cavity demonstrates a successful application of using the AMM for designing cavities with tailored modes for bespoke applications.

\section{Conclusion} \label{sec:Conclusion}

This paper has demonstrated the practical use of the azimuthally modulated method (AMM) in designing rf cavities for use in accelerator beamlines. The approach is underpinned by derivations for the change in both longitudinal  and transverse momentum of an ultra-relativistic, parallel particle traversing an azimuthally modulated cavity with beam pipes. We showed that the derived equations are in excellent agreement with simulation, even up to moderately non-relativistic speeds ($\gamma~\sim$~\qty{10}{}). Subsequently, we demonstrated two example applications: first, negating unwanted transverse multipoles in an accelerating mode, and second, combining transverse multipoles to create a bespoke multipolar field for uniformising a beam. 

The demonstration of a multipole-free accelerating structure is particularly interesting and warrants further investigation for application in injectors and precision machines. Another method of influencing the multipolar content of a mode is by shaping the iris of the beam pipe, and it would be valuable to develop a precise framework for controlling the multipoles in line with the AMM. Future research could also develop the AMM further by studying the effects of longitudinal asymmetries on the multipolar form of the change in momentum.

\section*{Acknowledgements}
The authors are very grateful to Andrea Latina for his support of this work, particularly his advice and assistance with \textsc{RF-Track} simulations; to Steinar Stapnes for his continual support on this project; and to Luke Dyks for his initial guidance on uniform beams. 

\section*{Data Availability}

The data are not publicly available. The data are available from the authors upon reasonable request.

\bibliography{apssamp}

@article{DecelerationCLIC,
  title = {{$X$}-band RF power production and deceleration in the two-beam test stand of the {Compact Linear Collider} test facility},
  author = {Adli, E. and Ruber, R. and Ziemann, V. and Corsini, R. and Dubrovskiy, A. and Syratchev, I.},
  journal = {Phys. Rev. ST Accel. Beams},
  volume = {14},
  issue = {8},
  pages = {081001},
  numpages = {11},
  year = {2011},
  month = {Aug},
  publisher = {American Physical Society},
  doi = {10.1103/PhysRevSTAB.14.081001},
  url = {https://link.aps.org/doi/10.1103/PhysRevSTAB.14.081001}
}

@article{DecelerationERL,
  title = {Ultimate energy recovery from spent relativistic electron beam in energy recovery linear accelerators},
  author = {Konoplev, I. V. and Shashkov, Ya. and Bulygin, A. and Gusarova, M. A. and Marhauser, F.},
  journal = {Phys. Rev. Accel. Beams},
  volume = {23},
  issue = {7},
  pages = {071601},
  numpages = {8},
  year = {2020},
  month = {Jul},
  publisher = {American Physical Society},
  doi = {10.1103/PhysRevAccelBeams.23.071601},
  url = {https://link.aps.org/doi/10.1103/PhysRevAccelBeams.23.071601}
}

@article{VelocityBunching,
  title = {Velocity bunching of high-brightness electron beams},
  author = {Anderson, S. G. and Musumeci, P. and Rosenzweig, J. B. and Brown, W. J. and England, R. J. and Ferrario, M. and Jacob, J. S. and Thompson, M. C. and Travish, G. and Tremaine, A. M. and Yoder, R.},
  journal = {Phys. Rev. ST Accel. Beams},
  volume = {8},
  issue = {1},
  pages = {014401},
  numpages = {22},
  year = {2005},
  month = {Jan},
  publisher = {American Physical Society},
  doi = {10.1103/PhysRevSTAB.8.014401},
  url = {https://link.aps.org/doi/10.1103/PhysRevSTAB.8.014401}
}

@article{BunchCompression,
    author = "Di Mitri, Simone",
    editor = "Bayley, R.",
    title = "{Bunch-length compressors}",
    doi = "10.23730/CYRSP-2018-001.363",
    journal = "CERN Yellow Rep. School Proc.",
    volume = "1",
    pages = "363",
    year = "2018"
}

@article{Linearisation1,
  title = {Tomography of longitudinal phase space linearization for the generation of attosecond electron bunches},
  author = {Schaap, B. H. and Musumeci, P.},
  journal = {Phys. Rev. Accel. Beams},
  volume = {28},
  issue = {1},
  pages = {012802},
  numpages = {8},
  year = {2025},
  month = {Jan},
  publisher = {American Physical Society},
  doi = {10.1103/PhysRevAccelBeams.28.012802},
  url = {https://link.aps.org/doi/10.1103/PhysRevAccelBeams.28.012802}
}

@article{Lineariser2,
  title = {Ka-band linearizer structFure studies for a compact light source},
  author = {Castilla, A. and Apsimon, R. and Burt, G. and Wu, X. and Latina, A. and Liu, X. and Syratchev, I. and Wuensch, W. and Spataro, B. and Cross, A. W.},
  journal = {Phys. Rev. Accel. Beams},
  volume = {25},
  issue = {11},
  pages = {112001},
  numpages = {13},
  year = {2022},
  month = {Nov},
  publisher = {American Physical Society},
  doi = {10.1103/PhysRevAccelBeams.25.112001},
  url = {https://link.aps.org/doi/10.1103/PhysRevAccelBeams.25.112001}
}

@article{EnergyCompression,
  title = {Radial focusing and energy compression of a laser-produced proton beam by a synchronous rf field},
  author = {Ikegami, M. and Nakamura, S. and Iwashita, Y. and Shirai, T. and Souda, H. and Tajima, Y. and Tanabe, M. and Tongu, H. and Itoh, H. and Shintaku, H. and others},
  journal = {Phys. Rev. ST Accel. Beams},
  volume = {12},
  issue = {6},
  pages = {063501},
  numpages = {6},
  year = {2009},
  month = {Jun},
  publisher = {American Physical Society},
  doi = {10.1103/PhysRevSTAB.12.063501},
  url = {https://link.aps.org/doi/10.1103/PhysRevSTAB.12.063501}
}

@techreport{KaonSeparator,
    author = "McAshan, M. and Wanzenberg, R.",
    title = "{RF} Design of a Transverse Mode Cavity for Kaon Separation",
    reportNumber = "FERMILAB-TM-2144",
    institution = "Fermi National Accelerator Laboratory Report No. FERMILAB-TM-2144",
    doi = "10.2172/780614",
    month = "{Mar}",
    year = "2001"
}

@article{DeflectorBrookhaven,
    author = {Hahn, H. and Halama, H. J.},
    title = {Design of the deflector for the rf beam separator at the {Brookhaven AGS}},
    journal = {Rev. Sci. Instrum.},
    volume = {36},
    number = {12},
    pages = {1788-1796},
    year = {1965},
    month = {Dec},
    issn = {0034-6748},
    doi = {10.1063/1.1719466},
    url = {https://doi.org/10.1063/1.1719466},
    eprint = {https://pubs.aip.org/aip/rsi/article-pdf/36/12/1788/19251643/1788\_1\_online.pdf},
}

@article{DeflectingTsinhua,
author = {Lin, X.-C. and Zha, H. and Shi, J. and Zhou, L.-Y. and Liu, S. and Gao, J. and Chen, H.-B.},
year = {2021},
month = {Apr},
pages = {36},
title = {Development of a seven-cell {$S$}-band standing-wave RF-deflecting cavity for {Tsinghua Thomson scattering X-ray} source},
volume = {32},
journal = {Nucl. Sci. Tech.},
doi = {10.1007/s41365-021-00871-5}
}

@article{Polarix,
  title = {Novel {$X$}-band transverse deflection structure with variable polarization},
  author = {Craievich, P. and Bopp, M. and Braun, H.-H. and Citterio, A. and Fortunati, R. and Ganter, R. and Kleeb, T. and Marcellini, F. and Pedrozzi, M. and Prat, E. and others},
  journal = {Phys. Rev. Accel. Beams},
  volume = {23},
  issue = {11},
  pages = {112001},
  numpages = {22},
  year = {2020},
  month = {Nov},
  publisher = {American Physical Society},
  doi = {10.1103/PhysRevAccelBeams.23.112001},
  url = {https://link.aps.org/doi/10.1103/PhysRevAccelBeams.23.112001}
}

@article{EmittanceExchange1,
  title = {Transverse-to-longitudinal emittance exchange to improve performance of high-gain free-electron lasers},
  author = {Emma, P. and Huang, Z. and Kim, K.-J. and Piot, P.},
  journal = {Phys. Rev. ST Accel. Beams},
  volume = {9},
  issue = {10},
  pages = {100702},
  numpages = {8},
  year = {2006},
  month = {Oct},
  publisher = {American Physical Society},
  doi = {10.1103/PhysRevSTAB.9.100702},
  url = {https://link.aps.org/doi/10.1103/PhysRevSTAB.9.100702}
}

@article{Emittance2,
  title = {Beam shaping using an ultrahigh vacuum multileaf collimator and emittance exchange beamline},
  author = {Majernik, N. and Andonian, G. and Lynn, W. and Kim, S. and Lorch, C. and Roussel, R. and Doran, S. and Wisniewski, E. and Whiteford, C. and Piot, P. and Power, J. and Rosenzweig, J. B.},
  journal = {Phys. Rev. Accel. Beams},
  volume = {26},
  issue = {2},
  pages = {022801},
  numpages = {8},
  year = {2023},
  month = {Feb},
  publisher = {American Physical Society},
  doi = {10.1103/PhysRevAccelBeams.26.022801},
  url = {https://link.aps.org/doi/10.1103/PhysRevAccelBeams.26.022801}
}

@article{ShortXray1,
  title = {Simulation and analysis of using deflecting cavities to produce short {X}-ray pulses with the Advanced Photon Source},
  author = {Borland, M.},
  journal = {Phys. Rev. ST Accel. Beams},
  volume = {8},
  issue = {7},
  pages = {074001},
  numpages = {18},
  year = {2005},
  month = {Jul},
  publisher = {American Physical Society},
  doi = {10.1103/PhysRevSTAB.8.074001},
  url = {https://link.aps.org/doi/10.1103/PhysRevSTAB.8.074001}
}

@article{ShortXray2,
doi = {10.1088/1748-0221/5/01/T01001},
url = {https://dx.doi.org/10.1088/1748-0221/5/01/T01001},
year = {2010},
month = {Jan},
journal = {J. Instrum.},
volume = {5},
number = {01},
pages = {T01001},
author = {H Ghasem and G H Luo and A Mohammadzadeh},
title = {Utilization of transverse deflecting RF cavities in the designed {QBA lattice of 3 GeV Taiwan Photon Source}},
}

@article{Crab1,
    author = "Palmer, R. B.",
    title = "Energy Scaling, Crab Crossing, and the Pair Problem",
    reportNumber = "SLAC-PUB-4707",
    journal = "eConf",
    volume = "C8806271",
    pages = "613--619",
    year = "1988"
}

@article{Crab2,
  title = {Coupled beam motion in a storage ring with crab cavities},
  author = {Huang, X.},
  journal = {Phys. Rev. Accel. Beams},
  volume = {19},
  issue = {2},
  pages = {024001},
  numpages = {11},
  year = {2016},
  month = {Feb},
  publisher = {American Physical Society},
  doi = {10.1103/PhysRevAccelBeams.19.024001},
  url = {https://link.aps.org/doi/10.1103/PhysRevAccelBeams.19.024001}
}

@article{LandauDamping,
  title = {Radio frequency quadrupole for {L}andau damping in accelerators},
  author = {Grudiev, A.},
  journal = {Phys. Rev. ST Accel. Beams},
  volume = {17},
  issue = {1},
  pages = {011001},
  numpages = {5},
  year = {2014},
  month = {Jan},
  publisher = {American Physical Society},
  doi = {10.1103/PhysRevSTAB.17.011001},
  url = {https://link.aps.org/doi/10.1103/PhysRevSTAB.17.011001}
}

@article{Landau2,
  title = {Analysis of transverse beam stabilization with radio frequency quadrupoles},
  author = {Schenk, M. and Grudiev, A. and Li, K. and Papke, K.},
  journal = {Phys. Rev. Accel. Beams},
  volume = {20},
  issue = {10},
  pages = {104402},
  numpages = {15},
  year = {2017},
  month = {Oct},
  publisher = {American Physical Society},
  doi = {10.1103/PhysRevAccelBeams.20.104402},
  url = {https://link.aps.org/doi/10.1103/PhysRevAccelBeams.20.104402}
}

@article{Wakefield1,
    author = "Wilson, P. B.",
    editor = "Month, M. and Dienes, M.",
    title = "Introduction to Wake Fields and Wake Potentials",
    reportNumber = "SLAC-PUB-4547, SLAC-AP-66, SLAC-AP-066",
    doi = "10.1063/1.38045",
    journal = "AIP Conf. Proc.",
    volume = "184",
    pages = "525--564",
    year = "1989"
}

@article{Wakefield2,
      author        = "Vaganian, S. and Henke, H.",
      title         = "The {Panofsky-Wenzel} Theorem and General Relations For The
                       wake Potential",
      journal       = "Part. Accel.",
      volume        = "48",
      pages         = "239-242",
      year          = "1995",
      url           = "https://cds.cern.ch/record/1108316",
}

@article{Wakefield3,
    author = "Dohlus, M. and Wanzenberg, W.",
    editor = "Herr, W.",
    title = "An Introduction to Wake Fields and Impedances",
    doi = "10.23730/CYRSP-2017-003.15",
    journal = "CERN Yellow Rep. School Proc.",
    volume = "3",
    pages = "15",
    year = "2017"
}

@article{EmittanceGrowth,
  title = {Emittance growth due to multipole transverse magnetic modes in an rf gun},
  author = {Chae, M. S. and Hong, J. H. and Parc, Y. W. and Ko, In Soo and Park, S. J. and Qian, H. J. and Huang, W. H. and Tang, C. X.},
  journal = {Phys. Rev. ST Accel. Beams},
  volume = {14},
  issue = {10},
  pages = {104203},
  numpages = {13},
  year = {2011},
  month = {Oct},
  publisher = {American Physical Society},
  doi = {10.1103/PhysRevSTAB.14.104203},
  url = {https://link.aps.org/doi/10.1103/PhysRevSTAB.14.104203}
}

@article{DualFeedCoupler,
  title = {Commissioning the {Linac Coherent Light Source} injector},
  author = {Akre, R. and Dowell, D. and Emma, P. and Frisch, J. and Gilevich, S. and Hays, G. and Hering, Ph. and Iverson, R. and Limborg-Deprey, C. and Loos, H. and Miahnahri, A. and Schmerge, J. and Turner, J. and Welch, J. and White, W. and Wu, J.},
  journal = {Phys. Rev. ST Accel. Beams},
  volume = {11},
  issue = {3},
  pages = {030703},
  numpages = {20},
  year = {2008},
  month = {Mar},
  publisher = {American Physical Society},
  doi = {10.1103/PhysRevSTAB.11.030703},
  url = {https://link.aps.org/doi/10.1103/PhysRevSTAB.11.030703}
}

@article{QuadFeed1,
doi = {10.1143/JJAP.49.086401},
url = {https://dx.doi.org/10.1143/JJAP.49.086401},
year = {2010},
month = {Aug},
publisher = {},
volume = {49},
number = {8R},
pages = {086401},
author = {Moon, S. and Hong, J. and Parc, Y. and Cho, M. and Namkung, W. and Ko, I. S. and Park, S.-J.},
title = {Reduction of Multipole Fields in Photocathode RF Gun},
journal = {Jpn. J. Appl. Phys.}
}

@article{QuadFeed2,
title = {Ultra-high brightness electron beams from very-high field cryogenic radiofrequency photocathode sources},
journal = {Nucl. Instrum. Methods Phys. Res., Sect. A},
volume = {909},
pages = {224-228},
year = {2018},
issn = {0168-9002},
doi = {https://doi.org/10.1016/j.nima.2018.01.061},
url = {https://www.sciencedirect.com/science/article/pii/S0168900218300780},
author = {J.B. Rosenzweig and A. Cahill and B. Carlsten and G. Castorina and M. Croia and C. Emma and A. Fukusawa and B. Spataro and D. Alesini and V. Dolgashev and M. Ferrario and G. Lawler and R. Li and 
C. Limborg and 
J. Maxson and 
P. Musumeci and
R. Pompili and 
S. Tantawi and
O. Williams}
}

@article{UniformIrradiation,
  title = {Tunable and compact permanent magnet uniform spreading beamline for electron irradiation accelerators},
  author = {Zhou, Z. and Sheng, K. and Zhang, Z. and Xiong, M. and Li, H. and Huang, J.},
  journal = {Phys. Rev. Accel. Beams},
  volume = {26},
  issue = {9},
  pages = {092402},
  numpages = {12},
  year = {2023},
  month = {Sep},
  publisher = {American Physical Society},
  doi = {10.1103/PhysRevAccelBeams.26.092402},
  url = {https://link.aps.org/doi/10.1103/PhysRevAccelBeams.26.092402}
}

@article{HEER, title={Generation of uniform transverse beam distributions for high-energy electron radiography}, volume={36}, DOI={10.1017/S0263034618000265}, number={3}, 
journal={Laser Part. Beams}, author={Zhao, Q.T. and Cao, S.C. and Cheng, R. and Du, Y.C. and Shen, X.K. and Wang, Y.R. and Xiao, J.H. and Zong, Y. and Zhu, Y.L. and Zhou, Y.W. and et al.}, year={2018}, pages={313–322}}

@inproceedings{Yongke,
  author       = {Zhao, Y. and Latina, A. and Doebert S.},
  title        = {Update of the {CLIC} positron source},
  booktitle    = {Proceedings of the International Workshop on Future Linear Colliders (LCWS2023)},
  year         = {2023},
  url          = {https://agenda.linearcollider.org/event/10134/contributions/54773/},
  organization = {Linear Collider Collaboration},
  note         = {Accessed: April 10, 2025}
}

@article{DEFT,
    author = "Rossi, C. and Malyzhenkov, A. and Grudiev, A. and Latina, A.  and Frisch, B.  and Granados, E.  and Syratchev, I.  and Cravero, J.  and Bauche, J.  and Angoletta, M.  and Brunner, O.  and Wang, P.  and Corsini, R.  and Doebert, S.  and Wuensch, W. ",
    title = "{The Deep Electron FLASH Therapy facility}",
    doi = "10.18429/JACoW-LINAC2024-WEYA002",
    journal = "JACoW",
    volume = "LINAC2024",
    pages = "WEYA002",
    year = "2024"
}

@book{Lee,
  author    = {S.Y. Lee},
  title     = {Accelerator Physics},
  publisher = {World Scientific},
  year      = {2018},
  edition   = {4th},
  isbn      = {978-9813232863},
}

@book{Wangler,
  author    = {T. Wangler},
  title     = {RF Linear Accelerators},
  publisher = {Wiley},
  year      = {2008},
  isbn      = {978-0471740736},
}

@article{AbellTransfer,
  title = {Numerical computation of high-order transfer maps for rf cavities},
  author = {Abell, D. T.},
  journal = {Phys. Rev. ST Accel. Beams},
  volume = {9},
  issue = {5},
  pages = {052001},
  numpages = {13},
  year = {2006},
  month = {May},
  publisher = {American Physical Society},
  doi = {10.1103/PhysRevSTAB.9.052001},
  url = {https://link.aps.org/doi/10.1103/PhysRevSTAB.9.052001}
}

@article{PanofskyWenzel,
    author = {Panofsky, W. K. H. and Wenzel, W. A.},
    title = {Some Considerations Concerning the Transverse Deflection of Charged Particles in Radio‐Frequency Fields},
    journal = {Rev. Sci. Instrum.},
    volume = {27},
    number = {11},
    pages = {967-967},
    year = {1956},
    month = {Nov},
    issn = {0034-6748},
    doi = {10.1063/1.1715427},
    url = {https://doi.org/10.1063/1.1715427},
    eprint = {https://pubs.aip.org/aip/rsi/article-pdf/27/11/967/19098135/967\_1\_online.pdf},
}

@article{WroeAMM,
  title = {Creating exact multipolar fields with azimuthally modulated rf cavities},
  author = {Wroe, L. M. and Sheehy, S. L. and Apsimon, R. J.},
  journal = {Phys. Rev. Accel. Beams},
  volume = {25},
  issue = {6},
  pages = {062001},
  numpages = {15},
  year = {2022},
  month = {Jun},
  publisher = {American Physical Society},
  doi = {10.1103/PhysRevAccelBeams.25.062001},
  url = {https://link.aps.org/doi/10.1103/PhysRevAccelBeams.25.062001}
}

@phdthesis{WroeThesis,
  edition = {},
  number = {},
  journal = {},
  pages = {},
  publisher = {University of Oxford},
  school = {University of Oxford},
  title = {Novel hybrid multipolar RF cavities for transverse beam manipulations},
  volume = {},
  author = {Wroe, L. M.},
  editor = {},
  year = {2022},
  series = {}
}

@misc{cst_mws,
  author       = {CST Studio Suite},
  title        = {{CST Microwave Studio}},
  year         = {2025},
  version      = {2022},
  note         = {Version 2022},
  howpublished = {\url{https://www.cst.com}},
  type         = {Computer software},
}

@misc{Mathematica,
  author = {Wolfram Research{,} Inc.},
  title = {Mathematica, {V}ersion 14.2},
  url = {https://www.wolfram.com/mathematica},
  note = {Champaign, IL, 2024}
}

@techreport{RF-Track,
    address = {Geneva, Switzerland},
    author = {Latina, A.},
    doi = {10.5281/zenodo.3887085},
    institution = {CERN},
    title = {{RF-Track} Reference Manual},
    year = {2024}
  }

@book{Lapostolle1970,
  title        = {Linear Accelerators},
  editor       = {P. M. Lapostolle and A. L. Septier},
  publisher    = {North Holland Publishing Company},
  year         = {1970},
  isbn         = {0720401569},
  address      = {Amsterdam, Netherlands}
}

@article{UniformBeam,
  title = {Uniformization of the transverse beam profile by means of nonlinear focusing method},
  author = {Yuri, Y. and Miyawaki, N. and Kamiya, T. and Yokota, W. and Arakawa, K. and Fukuda, M.},
  journal = {Phys. Rev. ST Accel. Beams},
  volume = {10},
  issue = {10},
  pages = {104001},
  numpages = {11},
  year = {2007},
  month = {Oct},
  publisher = {American Physical Society},
  doi = {10.1103/PhysRevSTAB.10.104001},
  url = {https://link.aps.org/doi/10.1103/PhysRevSTAB.10.104001}
}
\bibliographystyle{unsrt}
\end{document}